\documentclass[aip, jcp, reprint,citeautoscript,floatfix]{revtex4-1}
\usepackage{paralist}
\usepackage{graphicx}
\usepackage{amssymb}
\usepackage{amsbsy}
\usepackage{multirow}
\usepackage{mathrsfs}
\usepackage{amsmath}
\usepackage[T1]{fontenc}
\DeclareMathAlphabet{\mathscrbf}{OMS}{mdugm}{b}{n}
\usepackage{bm}
\usepackage{color}
\usepackage{epstopdf}
\usepackage[normalem]{ulem}
\usepackage{verbatim}
\usepackage[table]{xcolor}
\usepackage{float}
\restylefloat{table}
\usepackage{placeins}
\usepackage{booktabs}
\usepackage{hyperref}
\usepackage{array,url,kantlipsum}

\DeclareMathAlphabet{\mathscrbf}{OMS}{mdugm}{b}{n}

\newcommand{\kT}{k_B T}

\newcommand{\pR}{p_R}

\newcommand{\br}{{\bf r}}

\newcommand{\bG}{{\bf G}}

\newcommand{\bF}{{\bf F}}

\newcommand{\bff}{{\bf f}}

\newcommand{\bR}{{\bf R}}

\newcommand{\bphi}{{\boldsymbol \phi}}
\newcommand{\bb}{{\boldsymbol b}}
\newcommand{\bw}{{\boldsymbol w}}

\newcommand{\cG}{{\mathscrbf{G}}}

\newcommand{\etal}{{\it{et al.}}}
\begin{document}

\author{Svenja J. W\"{o}rner}
\author{Tristan Bereau}
\author{Kurt Kremer}
\author{Joseph F. Rudzinski}
\email{rudzinski@mpip-mainz.mpg.de}
\affiliation{Max Planck Institute for Polymer Research, 55128 Mainz, Germany}
%%%%%%%%%%%%%%%%%%%%%%%%%%%%%%%%%%%%%%%%%%%%%%%%%%%%%%%%%%%%%%%%%%%%%
%% The document title should be given as usual. Some journals require
%% a running title from the author: this should be supplied as an
%% optional argument to \title.
%%%%%%%%%%%%%%%%%%%%%%%%%%%%%%%%%%%%%%%%%%%%%%%%%%%%%%%%%%%%%%%%%%%%%

%%%%%%%%%%%%%%%%%%%%%%%%%%%%%%%%%%%%%%%%%%%%%%%%%%%%%%%%%%%%%%%%%%%%%
%% Some journals require a list of abbreviations or keywords to be
%% supplied. These should be set up here, and will be printed after
%% the title and author information, if needed.
%%%%%%%%%%%%%%%%%%%%%%%%%%%%%%%%%%%%%%%%%%%%%%%%%%%%%%%%%%%%%%%%%%%%%
%\abbreviations{coarse-graining, g-YBG}

%\begin{document}
\begin{abstract}
Coarse-grained (CG) models are often parametrized to reproduce one-dimensional structural correlation functions of an atomically-detailed model along the degrees of freedom governing each interaction potential.
While cross correlations between these degrees of freedom inform the optimal set of interaction parameters, the correlations generated from the higher-resolution simulations are often too complex to act as an accurate proxy for the CG correlations. 
Instead, the most popular methods determine the interaction parameters iteratively, while assuming that individual interactions are uncorrelated.
While these iterative methods have been validated for a wide range of systems, they also have disadvantages when parametrizing models for multi-component systems or when refining previously established models to better reproduce particular structural features.
In this work, we propose two distinct approaches for the direct (i.e., non-iterative) parametrization of a CG model by adjusting the high-resolution cross correlations of an atomistic model in order to more accurately reflect correlations that will be generated by the resulting CG model.
The derived models more accurately describe the low-order structural features of the underlying AA model, while necessarily generating inherently distinct cross correlations compared with the atomically-detailed reference model.
We demonstrate the proposed methods for a one-site-per-molecule representation of liquid water, where pairwise interactions are incapable of reproducing the true tetrahedral solvation structure.
We then investigate the precise role that distinct cross-correlation features play in determining the correct pair correlation functions, evaluating the importance of the placement of correlation features as well as the balance between features appearing in different solvation shells.
\end{abstract}

\title{
Direct route to reproducing pair distribution functions with coarse-grained models via transformed atomistic cross correlations
}
\maketitle
\section{Introduction}

Coarse-grained (CG) simulation models, which represent multiple atoms with a single CG site, allow efficient investigations relative to their atomistic counterparts while also providing insight into the essential driving forces for molecular processes.
Bottom-up CG models retain chemical specificity by reproducing target properties of a higher-resolution, e.g., all-atom (AA), reference model.
Ideally, the CG model should reproduce the Boltzmann distribution generated by the reference model, when viewed at the CG level of resolution.
The many-body potential of mean force (MB-PMF) is the high-dimensional free-energy function that achieves this goal when employed to model interactions between CG sites~\cite{Kirkwood:1935ys,Noid:2008a}.

In practice, the MB-PMF must be approximated, either explicitly (e.g., through a variational principle) or implicitly (e.g., by targeting lower-dimensional distribution functions).
For example, given a set of basis functions (i.e., a functional representation for interactions between CG sites), the multiscale coarse-graining (MS-CG) method~\cite{Izvekov:2005d,Izvekov:2005e,Noid:2008a} provides a variationally-optimal approximation to the MB-PMF.
This approach aims to determine the set of CG interaction parameters that will reproduce the total force on each CG site, for each configuration sampled by the reference model.
However, the method can also be recast in terms of matching structural distributions, through a connection to a generalized formulation of the well-established Yvon-Born-Green integral equation framework from liquid state theory~\cite{Mullinax:2009b,Mullinax:2010,Rudzinski:2015}.

For a basis set corresponding to pairwise interactions between nonbonded CG sites, the resulting generalized Yvon-Born-Green (g-YBG) equations provide a link between the interaction parameters and a set of two-body structural correlation functions~\cite{Ellis:2011a}---directly related to radial distribution functions (RDFs).
The interaction parameters and resulting structural correlation functions are related through a linear operator (i.e., a matrix) which quantifies the cross correlations between pairs of interaction types~\cite{Rudzinski:2012vn}.
In the limit of a complete basis set, e.g., when both the two-body structural correlation functions and the cross-correlation matrix are generated by the CG model, this relationship becomes exact.
To determine the set of parameters that will reproduce the two-body correlations generated by the reference model, the MS-CG method employs cross correlations generated by the higher-resolution reference model as a proxy for the correlations that will be generated by the resulting CG model~\cite{Rudzinski:2014}.
In the case that the CG model is able to reproduce the AA cross correlations, the MS-CG model will accurately model both the two-body correlation functions and also the cross correlations.
On the other hand, when the CG model is incapable of reproducing the AA cross correlations, due to basis set limitations, the MS-CG model will likely fail to reproduce either set of correlation functions.
Although alternative approaches based on the Ornstein-Zernike equation can alleviate these issues in some cases through implicit treatment of the cross correlations~\cite{Mashayak:2018}, the associated closure relations give rise to uncontrolled approximations which can result in significant errors as the complexity of the model increases.

In the absence of cross correlation information, one can assume that each basis function is independent from one another.
This approach, known as direct Boltzmann inversion (DBI)~\cite{Noid:2013,Noid:2013uq}, models pairwise interactions between CG sites with the pair potential of mean force.
In the context of the g-YBG equations, this approach corresponds to setting the three-body contributions of the cross-correlation matrix to zero~\cite{Rudzinski:2012vn}.
This approximation is typically only valid at very low densities or for long chain polymer melts, where entropic contributions overshadow the details of the nonbonded interactions~\cite{Tschop:1998lj}.
%, although it has been successfully employed to construct accurate models of polymer melts
In cases where the DBI model presents significant errors, one can iteratively refine the model to reproduce the RDFs via updates schemes which either (i) ignore cross correlations altogether (iterative Boltzmann inversion~\cite{Soper:1996ly,MullerPlathe:2002,Reith:2003tg}) or (ii) treat cross correlations approximately (inverse Monte Carlo~\cite{Lyubartsev:1995,Lyubartsev:1997}, iterative g-YBG~\cite{Cho:2009ve,Lu:2013uq,Rudzinski:2014}).

The iterative methods for matching distribution functions have been extremely useful for constructing bottom-up models of a wide range of soft matter systems~\cite{Karimi-Varzaneh:2012uq} including polymers~\cite{Qian:2008,Eslami:2011,Langeloth:2015}, liquid crystals~\cite{Peter:2008vc}, biological macromolecules~\cite{Savelyev:2009a,Savelyev:2010}, and ionic liquids~\cite{Savelyev:2009, Moradzadeh:2018}.
In the case of a pairwise set of basis functions, the latter class of methods can be directly linked to variational approaches for approximating the MB-PMF~\cite{Murtola:2009ys,Rudzinski:2011,Rudzinski:2014}.
Moreover, stochastic optimization techniques can significantly reduce the required number of iterations while increasing the robustness of the parametrization~\cite{Bilionis:2013vn,Shell:2016}.
Nevertheless, these methods can also present severe convergence problems~\cite{Jain:2006nx,Megariotis:2010fk,Fu:2012oq,Moore:2014}, even for the recovery of simple Lennard-Jones interaction potentials (i.e., no coarse-graining)~\cite{Rosenberger:2016}.  
%These issues can be sensitively dependent on chosen cut-off distances and enforced thermodynamic constraints~\cite{Rosenberger:2016}, and may be exacerbated as the complexity of the model increases due to (i) increasing the number of distinct interactions, (ii) enforcing model restrictions (e.g., fixing a subset of parameters based on a given reference model), or  (iii) incorporating reference data from multiple state points.
These issues can be sensitively dependent on chosen cut-off distances and enforced thermodynamic constraints~\cite{Rosenberger:2016}, and may be exacerbated as the complexity of the model increases due to increasing the number of distinct interactions or as the complexity of the optimization landscape increases through (i) enforcement of  model restrictions (e.g., fixing a subset of parameters based on a given reference model) or (ii) incorporation of reference data from multiple state points.
Moreover, structure-based potentials can be employed as a starting point for constructing CG models that also accurately represent thermodynamic~\cite{Das:2010,Dunn:2015,Dunn:2016} and dynamic~\cite{Izvekov:2006,Hijon:2010,Deichmann:2018} properties of the underlying reference model through a pressure-matching variational principle and the Mori-Zwanzig formalism, respectively.
This motivates the development of direct optimization methods that can be efficiently combined with these methodologies into a unified optimization procedure.

\begin{figure}[ht!]
\centering
\includegraphics[width=8.5cm]{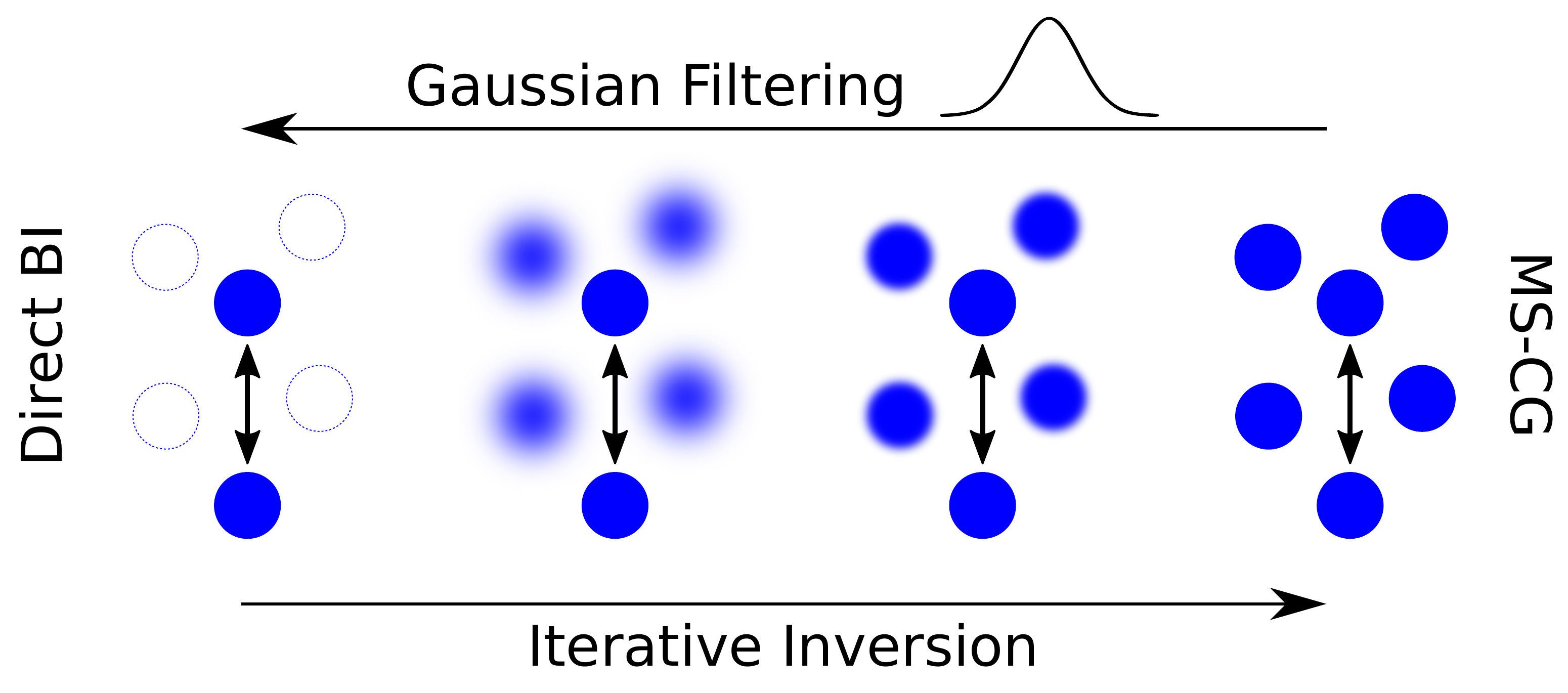}
\caption{
Schematic representation of altering the three-body contributions to the cross correlations of the high-resolution reference model, for the development of pairwise interaction potentials.
MS-CG and direct BI represent the extreme cases of incorporating the full high-resolution cross correlations and ignoring cross correlations, respectively.
The two methods proposed in this work---iterative inversion and Gaussian filtering---represent distinct interpolations between the extremes, which simplify the description of cross correlations to more accurately reflect the cross correlations that can be generated with the resulting CG model.
}
\label{fig:intro}
\end{figure}

In this work, we adapt the powerful framework provided by the MS-CG/g-YBG methods to match the practicality of the iterative methods for reproducing atomistic distribution functions.
We propose two complementary procedures to transform AA cross correlations to more accurately reflect correlations that can be modeled with a given set of basis functions for representing CG interactions.
The first method---referred to as iterative inversion---utilizes the natural structure of the AA cross-correlation matrix by performing a Neumann series expansion to obtain various orders of approximation to the full high-resolution correlations.
The second method---referred to as Gaussian filtering---softens the features of the AA cross-correlation matrix by applying an image blurring technique.
Both approaches effectively interpolate between employing the full set of AA cross correlations (MS-CG) and ignoring the three-body contributions to these cross correlations (DBI), see Figure~\ref{fig:intro}.
The methods are demonstrated on a one-site representation for liquid water~\cite{Praprotnik:2007bl,Wang:2009ol,Molinero:2009fu,Jochum:2012wa,Fritsch:2014,Hadley:2012}, where the tetrahedral solvation structure due to hydrogen-bonding interactions gives rise to cross correlations at the CG level of resolution that cannot be reproduced by a model employing standard pairwise interactions~\cite{Ruhle:2009wx,Chaimovich:2009tw}.
Although these discrepancies can be resolved by extending the basis set to include three-body potentials~\cite{Larini:2010dq,Das:2012c,Lu:2014jctc,Scherer:2018}, the pairwise case provides an ideal validation for the proposed approaches, where the MS-CG model fails to reproduce the RDF while the iterative methods match the RDF by disregarding higher-order structural correlations.

%Our analysis suggests that an accurate representation of the placement of correlation features (e.g., location of the correlation peak maxima) may be more important than quantitative reproduction of the correlations, motivating simple approaches for generating cross correlations given the CG representation.
Our analysis suggests that focusing on the accurate representation of the placement of correlation features (e.g., location of the correlation extrema) may be an advantageous approach for determining the optimal set of CG interaction parameters, motivating simple methods for generating cross correlations given the CG representation.
%Additionally, differences in the structural distributions between models are found to be dominantly caused by changes in the direct force, while changes in the cross correlations tend to have a relatively small effect.
Additionally, differences in the structural distributions between models are found to be dominated by contributions from the direct force, while changes in the cross correlations tend to have a comparatively smaller effect.
Differences in the direct force at larger distances are largely suppressed by the indirect contributions from the environment, in conjunction with Weeks-Chandler-Andersen theory~\cite{Weeks:1971}, although the impact of these forces can be significant depending on their precise form.
In particular, a balance between forces in the first and second solvation shells is crucial for reproducing the pair correlation functions of the reference model for liquid water, in agreement with previous work using core-softened potentials~\cite{Yan:2006,Yan:2008,Chaimovich:2009tw}.

The manuscript is organized as follows:
Section~\ref{sec:Theory} provides a brief overview of the g-YBG equations, describes properties of the corresponding cross-correlation matrix, and introduces the proposed transformations to be applied to the AA cross-correlation matrix.
Section~\ref{sec:methods} presents the computational details for the molecular simulations and force field calculations.
In Section~\ref{sec:results}, we describe the results of applying the proposed methods to a one-site representation of liquid water and discuss the general consequences of these results.
Finally, Section~\ref{sec:conc} summarizes the take-home messages and presents an outlook for future work.

\section{Theory}
\label{sec:Theory}

In this section all relevant aspects of the generalized Yvon-Born-Green (g-YBG) theory are introduced, along with the connection to the direct Boltzmann inversion (DBI) and multiscale coarse-graining (MS-CG) methods.
Then, two methods for simplifying the description of all-atom (AA) cross correlations, developed within the g-YBG framework, are presented.

\subsection{The G-YBG Framework}
\label{sec:mscg}

The g-YBG and, associated, MS-CG methods have been extensively described in previous work~\cite{Izvekov:2005d,Izvekov:2005e,Noid:2008a,Mullinax:2009b,Mullinax:2010,Noid:2013uq,Rudzinski:2015}.
Consider an AA reference model where each configuration $\br$ is represented by the Cartesian coordinates of $n$ atoms.
A set of mapping functions, $\{ M_{\bR I}(\br) \mid M_{\bR I}(\br) = \bR_I = \sum_{i\in I} c_i \br_i \}$, is chosen which systematically links an AA configuration to a coarser description, $\bR_I$, by representing groups of atoms with a single coarse-grained (CG) site, $I$.
The ``mapped AA ensemble'' can then be obtained by applying these mapping functions to each configuration of the AA model, e.g., in the canonical ensemble.
The many-body potential of mean force (MB-PMF) is the ideal potential for representing interactions at the CG level of resolution, such that the structural properties of the AA model will be quantitatively reproduced~\cite{Kirkwood:1935ys,Noid:2008a}.
The MB-PMF, which can be determined up to a configuration-independent constant, is given by
\begin{equation}
U^0(\bR) \propto -\kT \ln \pR(\bR) \,\, ,
\end{equation}
where $k_B$ is the Boltzmann constant, $T$ is the temperature, and $\pR(\bR)$ is the probability that an AA configuration maps to a particular CG configuration $\bR$.

The MB-PMF is a high-dimensional function which must be approximated by a simple set of CG interaction potentials.
To determine these interactions systematically, the approximate CG potential can be expressed as a linear combination of basis functions.
These basis functions may correspond to a particular functional form, e.g., a Lennard-Jones potential, or be completely flexible, e.g., spline functions.
For convenience, the remainder of the manuscript will consider the following special case: (i) the CG model represents each molecule with a single site and (ii) CG sites interact via a single nonbonded interaction, represented on a grid of piecewise constant functions in the space of force functions (described further below).
In this case, the CG potential is given by
\begin{equation}
U(\bR)= \sum^{\rm{pairs}}_{IJ} \sum_d \phi_{d} u_{d} (R_{IJ}) \,\, ,
\end{equation}
where the first sum runs over non-redundant pairs of distinct particles and $d$ represents the basis function index which encompasses a range of interparticle distances between sites $I$ and $J$.
$u_{d}$ is the $d^{th}$ potential energy basis function, centered at $R_d$ on the grid, and $\phi_{d}$ is the corresponding coefficient, which is to be optimized.

The force on site I in configuration $\bR$ is then given by
\begin{equation}
\bF_I(\bR) = - \frac{\partial U(\bR)}{\partial \bR_I} = \sum_{d} \phi_{d} \cG_{I,d}(\bR) \,\, ,
\end{equation} 
where
\begin{equation}
\cG_{I,d}(\bR) = \sum^{\rm{pairs}}_{IJ} \hat{\bR}_{IJ} f_{d}(R_{IJ}) \,\, ,
\end{equation}
$\hat{\bR}_{IJ}$ is the unit vector from $J$ to $I$, indicating the direction of the force acting on site $I$ from the interaction with site $J$, and $f_{d}(R_{IJ}) = -{\rm d}u_{d}(R_{IJ})/{\rm d} R_{IJ}$ is the corresponding force basis function.

The MS-CG method (commonly referred to as ``force-matching'') approximates the MB-PMF by projecting the many-body mean force, i.e., $\bF^0(\bR) = - \nabla U^0(\bR)$, onto the set of basis vectors $\{ \cG_{d}(\bR) \}$.
It can be demonstrated that this projection corresponds to a linear least squares problem, with the optimization functional~\cite{Noid:2008a,Noid:2008b}
\begin{equation}
\chi^2 [\bF] =\frac{1}{3N} \bigg\langle \sum_I |\bff_I(\br)-\bF_I(\bR)|^2 \bigg\rangle \,\, ,
\end{equation}
where $\bF=\{\bF_1(\bR),\bF_2(\bR), ..., \bF_N(\bR)\}$ represents a trial force field, $N$ is the number of CG sites, $\bff_I(\br)$ is the total force mapped onto site $I$ from the AA model in configuration $\br$, and $\langle \cdot \rangle$ is an ensemble average.
In other words, the MS-CG method performs a force-matching procedure by varying the parameters $\bphi$ of the CG model to best reproduce the mapped AA forces.

Equivalently, the problem can be expressed in terms of a set of linear equations~\cite{Noid:2008b}:
\begin{equation}
b_d = \sum_{d'} G_{dd'} \phi_{d'} \,\, .
\label{eq:setlineq}
\end{equation}
$\bb$ is a vector of force correlation functions which correspond to projections of $F^0$ onto the each of the basis vectors, i.e., $b_d = \cG_d \odot F^0$, where $\odot$ represents an inner product between force fields.
$G_{dd'}$ quantifies the ``angle'' formed by pairs of basis vectors: $G_{dd'} = \cG_d \odot \cG_d'$.

Equation~\ref{eq:setlineq} can be transformed to depend only on structural information, revealing the set of equations as a generalization of the Yvon-Born-Green integral equation framework from liquid state theory~\cite{Mullinax:2009b,Mullinax:2010}.
Within this formulation, $\bb$ corresponds to a structural correlation function that is directly related to the radial distribution function (RDF):
\begin{equation}
b_d = k_B T c R_d^2 \left (\frac{{\rm d} {\bm g}}{{\rm d} R} \right )_{\!\!d} \,\, ,
\label{eq:bd}
\end{equation}
where $c=(4\pi N)/(3 V)$ and $\bm g$ is the discretization of the RDF implied by the basis function representation.
$\left ({\rm d} {\bm g} / {\rm d} R \right )$ is meant as a numerical derivative of $\bm g$ with respect to interparticle distance $R$, given by the basis function centers $\{R_d\}$.

The correlation matrix $\bG$ also has a clear physical interpretation~\cite{Rudzinski:2012vn}.
First, it is useful to decompose $\bG$ into two matrices which, through Equation~\ref{eq:setlineq}, determine the direct and indirect contributions to $\bb$:
\begin{equation}
G_{dd'}=\bar{g}_d \delta_{dd'} + \bar{G}_{dd'} \,\, ,
\label{eq:decompose}
\end{equation} 
where $\delta_{dd'}$ is the Kronecker delta function.
The direct contribution $\bar{\bm g}$ is a correlation function that is again related to the RDF: $\bar{g_d} = c R_d g_d$.
%$\bar{g}(R) = c R^2 g(R)$, where we have switched from vector to function notation for simplicity.
%
$\bar{\bG}$, on the other hand, quantifies the cross correlations between pairs of interactions.
For non-bonded interactions, $\bar{\bG}$ characterizes the average angle formed between triplets of CG sites:
\begin{equation}
\bar{G}_{dd'} = \frac{1}{3N} \left \langle \sum^{{\rm triplets}}_{IJK} \cos \theta_{IJK} f_{d}(R_{IJ}) f_{d'}(R_{IK}) \right \rangle \,\, .
\label{eq:barG}
\end{equation}
For piecewise constant basis functions, $f_{d}(R_{IJ}) = 1$ if $R_d-0.5d\!R \leq R_{IJ} < R_d+0.5d\!R$ and 0 otherwise, where $d\!R$ represents the grid spacing.
$\bar{G}_{dd'}$ then corresponds to the ensemble average of the sum of cosines of the angle between triplets, represented on a 2D grid of $IJ$ and $IK$ pair distances.
This is schematically shown in the inset in Figure~\ref{fig:overview}(c).

Equation~\ref{eq:bd} clearly implies a relationship between $b(R)$ and the pair mean force, $-w'(R) = -\frac{\rm d}{{\rm d}R} [-\kT \ln  g(R) ]$.
Thus, using Equation~\ref{eq:decompose}, the pair mean force can be decomposed into direct and indirect contributions:
\begin{equation}
-w'_d = \frac{b_d}{\bar{g}_d} = \phi_d + \frac{1}{\bar{g}_d}  \sum_{d'} \bar{G}_{dd'} \phi_{d'} \,\, .
\label{eq:ppmf}
\end{equation}

The g-YBG equations (Equation~\ref{eq:setlineq}) represent an exact relationship between the force field parameters $\bphi$ and the structural correlation functions $\bb(\bphi)$, determined from molecular simulations, via the cross correlations, $\bG(\bphi)$, generated by the same model $\bphi$.
In contrast, the MS-CG method employs AA simulation data to calculate the correlation functions $\bb^{\rm AA}$ and $\bG^{\rm AA}$.
In other words, the approach attempts to predict the force field parameters $\bphi$ that will reproduce $\bb^{\rm AA}$, using $\bG^{\rm AA}$ as a proxy for the cross correlations of the CG model~\cite{Rudzinski:2014,Rudzinski:2014b}.
When $\bG^{\rm AA} \approx \bG(\bphi)$, the resulting model will approximately reproduce both sets of structural correlation functions, i.e., both two- and three-body correlations.
However, in practice the basis set employed to model interactions between CG sites is often too limited to accurately describe AA cross correlations, such that the resulting model demonstrates errors in $\bb$ and, consequently, also in the RDF.

The g-YBG framework provides a platform for iteratively refining the force field parameters to quantitatively reproduce the RDF, by solving the g-YBG equations self-consistently~\cite{Cho:2009ve,Lu:2013uq,Rudzinski:2014}.
In this approach, the cross correlations generated at each step are used in Equation~\ref{eq:setlineq} to obtain a new set of parameters that more accurately reproduce $\bb^{\rm AA}$.
Similarly, iterative Boltzmann inversion iteratively refines the CG parameters while assuming no cross correlations between the degrees of freedom governing interactions between CG sites (i.e., setting $\bar{\bG}$ to zero).
Notice that setting $\bar{\bG}$ to zero in Equation~\ref{eq:setlineq} corresponds to DBI.
Thus, the MS-CG method combined with the g-YBG framework encompasses an entire range of structure-based methods from the employment of AA reference cross correlations to the assumption of no cross correlations.

\subsection{Iterative Inversion}
\label{sec:theory_ii}

As shown in equation \ref{eq:decompose}, it is possible to decompose the g-YBG correlation matrix {\bf G} into direct and indirect contributions.
To simplify the notation in this section, we will denote the direct contribution matrix as $\bG_2$ and the indirect contribution matrix as $\bG_3$.
The decomposition of the correlation matrix can therefore be expanded as
\begin{eqnarray}
\bG &=& \bG_2 + \bG_3 \nonumber \\
    &=& \bG_2 + \bG_2 \bG_2^{-1} \bG_3 \nonumber \\ 
    &=& \bG_2 ({\bf I}  + \bG_2^{-1} \bG_3) \nonumber \\
    &=& \bG_2 \left( {\bf I} - (-\bG_2^{-1} \bG_3) \right) \,\, ,
\end{eqnarray}
with ${\bf I}$ being the identity matrix. Inverting this expression leads to
\begin{equation}
 \bG^{-1} = \left( {\bf I} - (-\bG_2^{-1} \bG_3)\right)^{-1} \bG_2^{-1} \,\, .
 \label{eq:Ginverse}
\end{equation}
The first term in equation \ref{eq:Ginverse} can be approximated using a Neumann series expansion~\cite{Suhubi:2003}:
\begin{equation}
 \bG^{-1}_{n} = \sum_{i=0}^{n} (-\bG_{2}^{-1} \bG_{3})^i \bG_{2}^{-1} \,\, .
 \label{eq:iterinversion}
\end{equation}
Note that employing the $0^{th}$ order term in the expansion to solve Equation~\ref{eq:setlineq} corresponds to DBI, since no indirect contributions are included.
By \emph{iteratively} calculating the inverse of $\bG$ via this expansion, various correlation matrices corresponding to an interpolation between the correlations used to determine the DBI and MS-CG models are recovered.

\subsection{Gaussian Filtering}
\label{sec:theory_gauss}
As an alternative approach to the iterative inversion method, we consider applying a Gaussian filter to the indirect contributions of the cross-correlation matrix, $\bar{\bG}$.
The filtered matrix is obtained using the following transformation:
\begin{equation}
\bar{G}^{\rm GF}_{ij}=\frac{1}{2\pi\sigma^2} \sum^{N_d}_{i'=1}\sum^{N_d}_{j'=1}\exp{\left(-\frac{(i'-i)^2+(j'-j)^2}{2\sigma^2}\right)} \bar{G}_{i'j'} \,\, ,
\label{eq:gaussfilter}
\end{equation}
where $\sigma$ is the standard deviation of the Gaussian function and $N_d$ is the dimension of the matrix, given by the basis function representation.

%%%%%%%%%%%%%%%%%%%%%%%%%%%%%%%%%%%%%%%%%%%%%%%%%%%%%%%%%%%%%%%%%%%%%%%%%5
\section{Methods}
\label{sec:methods}

\noindent \uline{All-atom (AA) simulations:}
The atomistic simulations of 1000 SPC/E\cite{Berendsen:1987} water molecules with periodic boundary conditions were previously performed by Scherer et al.\cite{Scherer:2018}, using version 5.1 of the GROMACS\cite{Abraham:2015} package.
The water molecules were equilibrated for 10 ns in the $NPT$ ensemble at a temperature of 300~K and a pressure of 1~bar, using the Berendsen barostat\cite{Berendsen:1984} with a time constant of 1~ps and using 4.5$\times 10^{-5}$ bar$^{-1}$ as the compressibility parameter for water.
For the subsequent production run, the water molecules were simulated for 10~ns in the $NVT$ ensemble at a density of 0.998~g~cm$^{-3}$. 
The stochastic dynamics algorithm\cite{VanGunsteren:1988} was used to integrate the equations of motion with a time step of 1~fs and a temperature coupling constant of 1~ps.
The smooth particle mesh Ewald method\cite{Essmann:1995} was used to treat the electrostatic interactions using cubic interpolation, a grid spacing of 0.12~nm and an Ewald accuracy parameter of 10$^{-5}$.
The cutoff for the Van der Waals interactions was set to 1.2~nm and the long-range dispersion correction was used for energy and pressure. 

\noindent \uline{Coarse-grained (CG) force fields:}
Each water molecule was represented by a single CG site, positioned at the molecular center of mass.
The iterative Boltzmann inversion (IBI) potential was obtained with the procedure described by R{\"u}hle \etal~\cite{Ruhle:2009wx}, using version 1.5 of the VOTCA toolkit~\cite{Ruhle:2009wx,Mashayak:2015}.
All other coarse-grained models were developed with the BOCS package~\cite{Dunn:2018}, which is available on GitHub.
A single nonbonded potential was employed to model interactions between CG sites.
The correlation functions in Equation~\ref{eq:setlineq} were evaluated between 0.22 and 1.2~nm using the ``delta'' basis set (i.e., piecewise constant functions) with a grid spacing of 0.002~nm.
We modified the equations using a trimming parameter of 0.01 and regularization parameter of 0.01~\cite{Dunn:2018}.
The set of linear equations were then solved using singular value decomposition after applying right-left preconditioning to make the equations dimensionless.

We also solved Equation~\ref{eq:setlineq} with an analytic power series basis with exponent values of 6 and 12, similar to a standard Lennard-Jones potential.
The resulting potential was then transformed to the Weeks-Chandler-Andersen (WCA) form by shifting the potential minima to 0 and then truncating the potential at larger distances.

\noindent \uline{CG simulations:}
All coarse-grained force fields were simulated using version 4.5.3 of the GROMACS package~\cite{Hess:2008}.
A characteristic length unit, $\mathscr{L}$, is defined as the diameter of one water molecule in the AA model (2 \AA).
Similar, the energy unit is defined as $\mathscr{E}=k_BT=1.38 \times 10^{-23}~$J~K$^{-1} \times 298~$K, and the mass of the CG bead as $\mathscr{M}=18 \times 1.66 \times 10^{-27}$~kg.
Then, the natural time unit of the simulation can be calculated as $\tau = \mathscr{L}\sqrt{\mathscr{M}/\mathscr{E}} \approx 0.54$~ps.
For simplicity, we let $\tau = 0.5$~ps.
%Starting from an AA configuration mapped to the CG representation, 1000 CG water sites were simulated for 10.5~ns in the $NVT$ ensemble at 298~K, using the stochastic dynamics integrator with a temperature coupling constant of 0.5~ps, a 1~fs time step, and periodic boundary conditions.
%The first 500~ps of each trajectory was removed for equilibration and the remaining 10~ns was used for analysis.
% tau = 0.54
%Starting from an AA configuration mapped to the CG representation, 1000 CG water sites were simulated for $19.444 \times 10^{3}~\tau$ in the $NVT$ ensemble at 298~K, using the stochastic dynamics integrator with a temperature coupling constant of 0.93~$\tau$, a $1.85 \times 10^{-3}~\tau$ time step, and periodic boundary conditions.
%The first 926~$\tau$ of each trajectory was removed for equilibration and the remaining $18.518 \times 10^{3}~\tau$ was used for analysis.
% tau = 0.50
Starting from an AA configuration mapped to the CG representation, 1000 CG water sites were simulated for $21 \times 10^{3}~\tau$ in the $NVT$ ensemble at 298~K, using the stochastic dynamics integrator with a temperature coupling constant of 1~$\tau$, a $2 \times 10^{-3}~\tau$ time step, and periodic boundary conditions.
The first 1000~$\tau$ of each trajectory was removed for equilibration and the remaining $2 \times 10^{4}~\tau$ was used for analysis.

\noindent \uline{Modifications of the correlation matrix:}
Iterative inversion via a Neumann series expansion was achieved with the \texttt{Iterative\_Inversion} directive in BOCS~\cite{Dunn:2018}.
The Gaussian filter was applied using the \texttt{gaussian\_filter()} 
function of the SciPy Multidimensional Image processing package \texttt{ndimage}~\cite{SciPy}. 
The array borders were handled with the constant mode, meaning all values outside the edges were set to 0. 
This allows the array to converge to zero at an infinite standard deviation.
The standard deviation is defined in multiples of bin sizes and varies between 0.02 and 0.08~nm for the different models.

%%%%%%%%%%%%%%%%%%%%%%%%%%%%%%%%%%%%%%%%%%%%%%%%%%%%%%%%%%%%%%%%%%%%%%%%%%%%%%%%%%%%%%%%%%%%%%%%%%%%%%%%%%%%%%%%%%%%%%%%%%%%%%%%%%%%%%%%%%%%%%%%%%%%%%%%%%%%%%%%%%%%%%%%%%%%%%%%%%%%%%%%%%%%%%%%%%%%%%%%%%%%%%%%%%%%%%%%%%%%%%%%%
\section{Results and discussion}
\label{sec:results}

Various bottom-up methods for coarse-graining take different approaches to reproduce structural correlation functions of the given reference system, e.g., obtained from simulations of an all-atom (AA) model.
For example, direct Boltzmann inversion (DBI) assumes that individual interactions are uncorrelated, an assumption that is typically far too crude in the condensed phase~\cite{Noid:2013,Noid:2013uq}.
In contrast, the multiscale coarse-graining (MS-CG) method uses cross correlations generated by the AA reference model to directly predict the optimal set of parameters that will result in the reproduction of the average force on each coarse-grained (CG) site~\cite{Izvekov:2005d,Izvekov:2005e,Noid:2008a}.
In the case that the CG model is capable of reproducing the AA cross correlations, the MS-CG framework guarantees the reproduction of both two- and three-body correlation functions along the degrees of freedom governing interactions in the CG force field through its link with the generalized Yvon-Born-Green (g-YBG) integral equation~\cite{Mullinax:2009b,Mullinax:2010,Rudzinski:2015}.
However, the molecular mechanics potentials typically employed to describe CG interactions can severely limit the model's ability to reproduce complex AA correlations~\cite{Rudzinski:2014,Rudzinski:2014b,Bereau:2018}.

A one-site CG representation for liquid water, extensively studied in previous works~\cite{Ruhle:2009wx,Larini:2010dq,Das:2012c,Lu:2014jctc,Scherer:2018}, provides a prototypical example.
The tetrahedral solvation structure gives rise to distinct cross correlations that cannot be reproduced by a CG model employing standard pairwise interactions.
As a consequence, this is an ideal test case for investigating the transformation of AA cross correlations to better reflect the correlations that can be generated by the CG model.
In the following, we apply two different approaches for the transformation---iterative inversion and Gaussian filtering---which effectively interpolate between the two extreme cases: DBI and MS-CG.

\subsection{DBI and MS-CG}

\begin{figure}[ht!]
\centering
    \includegraphics[width=8.5cm]{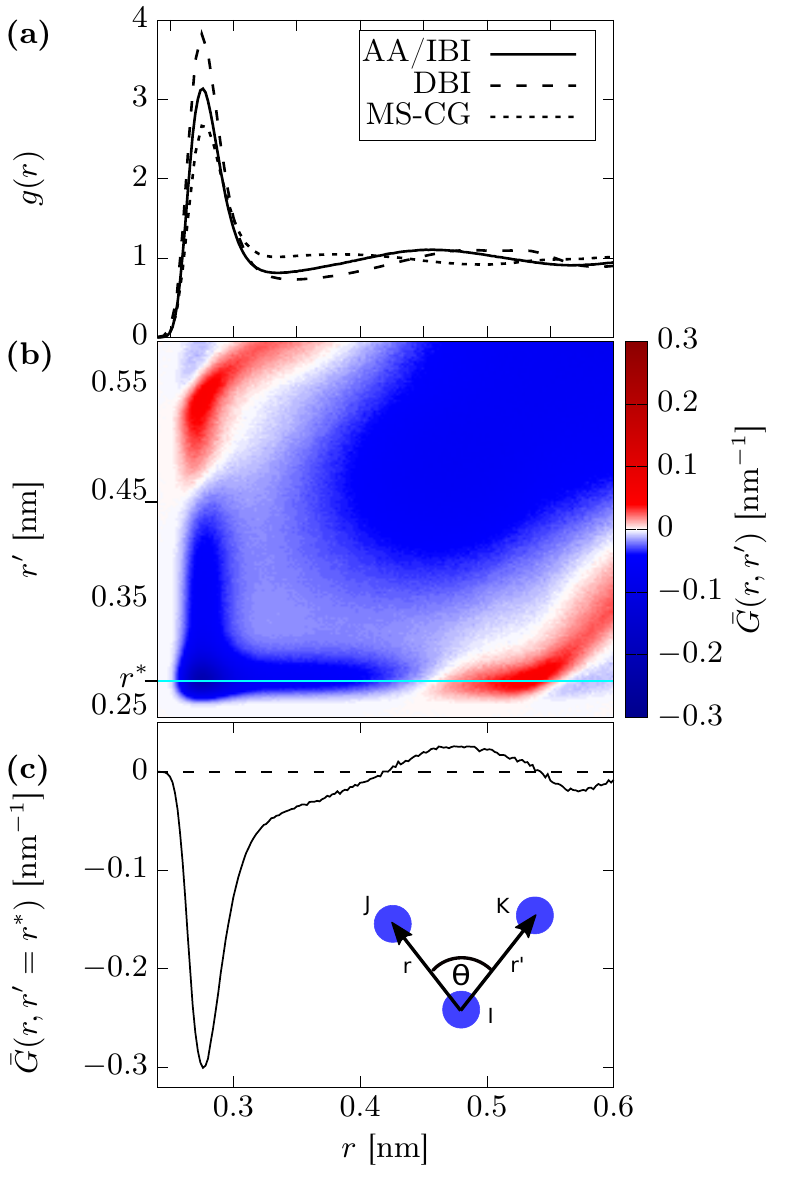}
  \caption{
(a) RDFs generated from simulations of the AA (solid curve), IBI (identical to AA curve by construction), DBI (long dashed curve), and MS-CG (short dashed curve) models.
(b) Three-body contributions to the MS-CG/g-YBG cross-correlation matrix (Equation~\ref{eq:barG}) generated by AA simulations, mapped to the CG representation. 
(c) Cross-section of cross-correlation matrix in panel (b), at fixed $r' = r^\ast = $ 0.276~nm, as indicated by the cyan line in panel b.  
The schematic inset shows a pair of particles, J and K, at distances $r$ and $r^\ast$ from the central particle, I. 
The average angle between all triplets with these distances constitutes the entry in the cross-correlation matrix $\bar{G}$ in panel (b). 
In panels (b) and (c) the entries of the cross-correlation matrix were multiplied by a factor of 100. 
}
  \label{fig:overview}
\end{figure}

In the context of employing cross correlations to determine the optimal CG potentials for reproducing structural features of an AA reference model, DBI and MS-CG can be viewed as two limiting cases---assuming no cross correlations and assuming that the cross correlations generated by the AA model are a good proxy for the correlations that will be generated by the resulting CG model, respectively.
Figure~\ref{fig:overview}(a) presents the radial distribution functions (RDFs) generated by the DBI (long dashed curve) and MS-CG (short dashed curve) models, compared with the target atomistic RDF (solid curve).
By assuming no correlations, the DBI model employs the pair potential of mean force, $w(r) = -\kT \ln g(r)$, to model interactions between CG sites.
As expected for a system in the condensed phase, this potential inherits features of the solvation structure, resulting in an overestimation of the attractive forces needed to stabilize the target structure, as indicated by the overstructuring of the first peak in the RDF.
In contrast, the MS-CG model underestimates the magnitude of the first peak in the RDF, indicating that the AA cross correlations are too strong on average, resulting in an underestimation of the attractive forces needed to stabilize the target structure.
This result also indicates that the IBI model, which reproduces the target RDF by construction, necessarily generates distinct cross correlations with respect to the AA model.
 
Figure~\ref{fig:overview}(b) presents the three-body contributions to the cross-correlation matrix, $\bar{G}$, generated by the AA model.
This matrix characterizes the correlation between a triplet of CG sites as the average angle between the sites when one pair is at a distance $r$ and the second pair is at a distance $r'$, as indicated by equation \ref{eq:barG} and schematically shown in the inset in Figure~\ref{fig:overview}(c).
Positive (negative) values indicate a preference for acute (obtuse) angles.
The dominant features of the matrix in Figure~\ref{fig:overview}(b) are generic to liquids~\cite{Rudzinski:2012vn}.
When $r \approx r'$, all angles between the triplet are possible with the exception of angles around 0~degrees, prevented by the excluded volume of the sites.
This feature is exaggerated at short distances, especially within the first solvation peak of the RDF, $r\approx r' \approx$ 0.276~nm in this case.
The positive strip at $r \approx r' +$ 0.25~nm is indicative of the solvation shell structure of the liquid, where each CG site falls within the first solvation shell of another site in the triplet.
Figure~\ref{fig:overview}(c) presents a cross-section of the matrix, indicated by the cyan line in panel (b).
This cross-section represents the correlation between three sites when the distance between one pair is kept fixed at the position of the first solvation shell peak in the RDF ($r' = r^* =$ 0.276 nm), while the distance between the second pair is varied.
The first correlation extremum characterizes the excluded volume effect as described above.
The second extremum is due to the fluctuating solvation structure of the liquid and, in particular, the tetrahedral ordering of the water molecules.
Note that DBI corresponds to setting all elements of $\bar{G}$ to zero (long dashed line in Figure~\ref{fig:overview}(c)) and then solving Equation~\ref{eq:setlineq}, while the MS-CG procedure employs $\bar{G}$ obtained from simulations of the AA model (black line in Figure~\ref{fig:overview}(c)).

\subsection{Iterative Inversion}

\begin{figure}[ht!]
\centering
    \includegraphics[width=8.5cm]{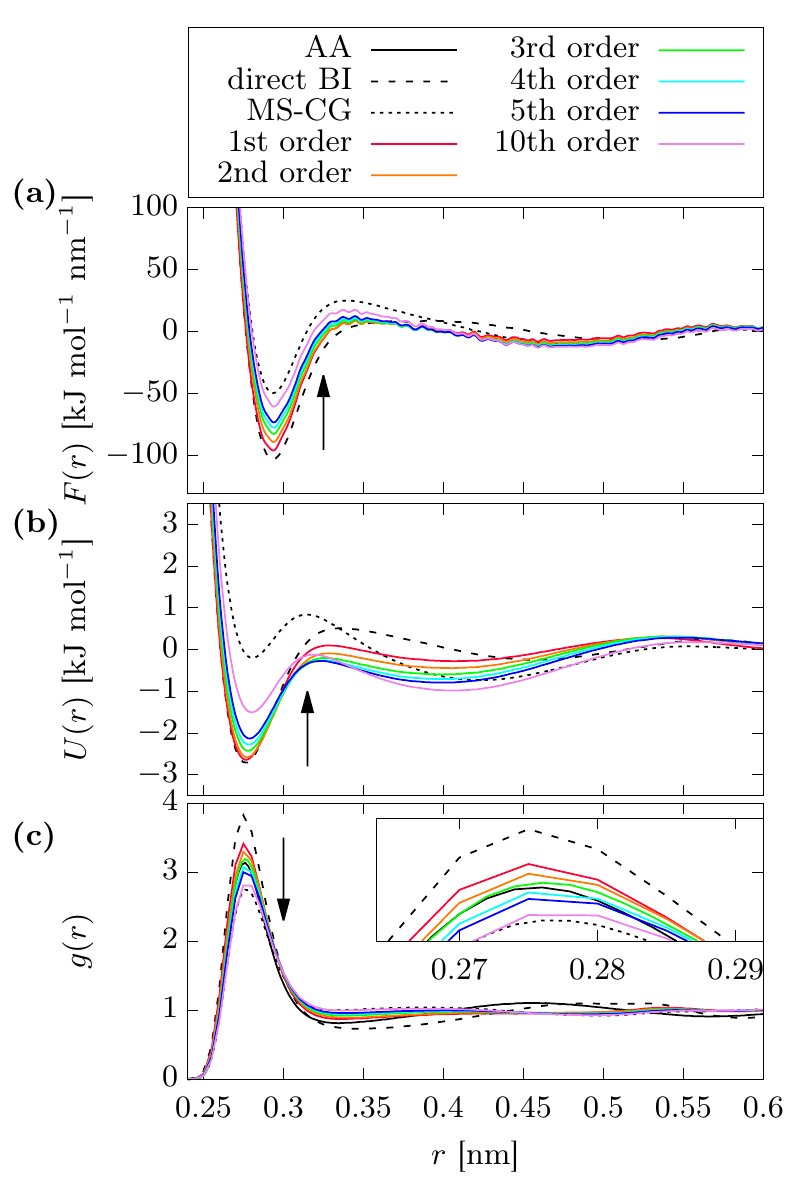}
  \caption{
Forces (a), potentials (b), and RDFs (c) corresponding to the DBI (long dashed curve) and MS-CG (short dashed curve) models as well as several models obtained from truncating the expansion of the inverse of the AA cross-correlation matrix at various orders, ranging from 1 to 10 (solid colored curves).
The AA RDF is presented as the solid black curve in panel (c).
Note that the $0^{th}$ order expansion coincides with DBI.
}
    \label{fig:iterinv}
\end{figure}

When solving the g-YBG equations (Equation~\ref{eq:setlineq}) for the force field $\bphi$, the inverse of the correlation matrix $\bG$ can be approximated by performing a Neumann series expansion, as described in Section~\ref{sec:theory_ii}. %, and then truncating the series after a finite number of terms.
Truncating the series after a finite number of terms (see Equation~\ref{eq:iterinversion}) allows one to systematically vary the degree to which cross correlations are taken into account. 
In contrast to the standard iterative techniques for coarse-graining (e.g., IBI), here ``iterative'' refers to the iterative solution to inverting the correlation matrix, and does not involve simulations of the CG model.
Panels (a) and (b) of Figure~\ref{fig:iterinv} present the forces and corresponding potentials, respectively, obtained from truncating the expansion of the inverse of the AA cross-correlation matrix at various orders, ranging from 1 to 10 (solid colored curves).
Note that the $0^{\rm \it th}$ order expansion model coincides with DBI.
Panel (a) demonstrates that the g-YBG iterative inversion method provides a smooth interpolation between the DBI (long dashed black curve) and MS-CG (thin dashed black curve) solutions.

Figure~\ref{fig:iterinv}(c) presents the RDFs generated by the iterative inversion models, compared with those generated from the AA (solid black curve), DBI, and MS-CG models.
The systematic decrease in the attractive force between sites at $r \approx$ 0.276~nm with increasing order of the expansion (panel (a)) leads to a corresponding increase in the height of the first solvation shell peak in the RDF.
Thus, we can identify the model which most closely matches the AA RDF, the $3^{\rm \it rd}$ order model in this case.
However, even for the 1$^{\rm \it st}$ order model, which incorporates the least amount of cross-correlation information, both the interactions and the resulting RDFs closely resemble those of the MS-CG model within the region of the second solvation shell, $r \approx$ 0.48~nm.
This result indicates that there exists dominant features of the AA cross-correlation matrix, likely predominantly associated with the second solvation shell region, which cannot be removed through the truncation of the iterative inversion.
Since these correlation features are responsible for systematic errors in the CG models, as indicated by the discrepancies in the second solvation peaks in panel (c) and discussed in more detail below, an alternative method is required in this case to simplify the AA cross correlations to better represent those attainable with the CG model.

\subsection{Gaussian Filtering}

\begin{figure}[ht!]
    \includegraphics[width=8.5cm]{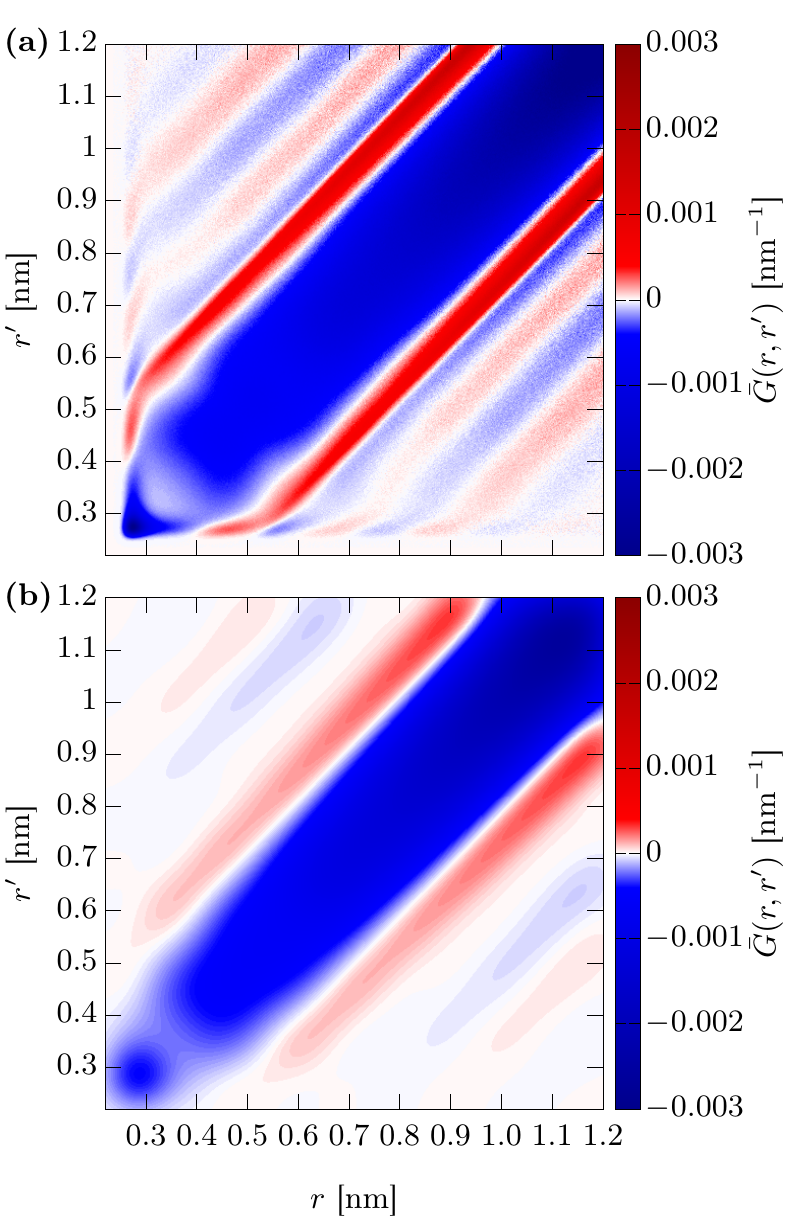}
\caption{
(a) Three-body contributions to the cross-correlation matrix determined from simulations of the AA model, mapped to the CG representation.
(b) Three-body contributions to the cross-correlation matrix determined by smoothing the AA cross-correlations in panel (a) using a Gaussian filter with a standard deviation of $\sigma$ = 0.04~nm.
}
\label{fig:matrices}
\end{figure}

Since the AA correlations appear to be too strong on average, a method is required to transform the cross-correlation matrix to reduce the overall magnitude of the correlations.
Motivated by the general idea that coarse-graining results in a smoothing of the free-energy landscape of the underlying system, we propose a Gaussian filter transformation, analogous to blurring the pixels of an image.
Rudzinski and Noid~\cite{Rudzinski:2014b} previously demonstrated the usefulness of such a transformation for simplifying AA cross correlations in order to construct structurally accurate, minimal models for helix-coil transitions.
Panels (a) and (b) of Figure~\ref{fig:matrices} present the AA cross-correlation matrix, $\bar{G}$, before and after the application of a Gaussian filter with a standard deviation of $\sigma$ = 0.04~nm, respectively.
In the limit of an infinitely large standard deviation, the entire matrix would be equal to zero, corresponding to DBI.

\begin{figure}[ht!]
    \includegraphics[width=8.5cm]{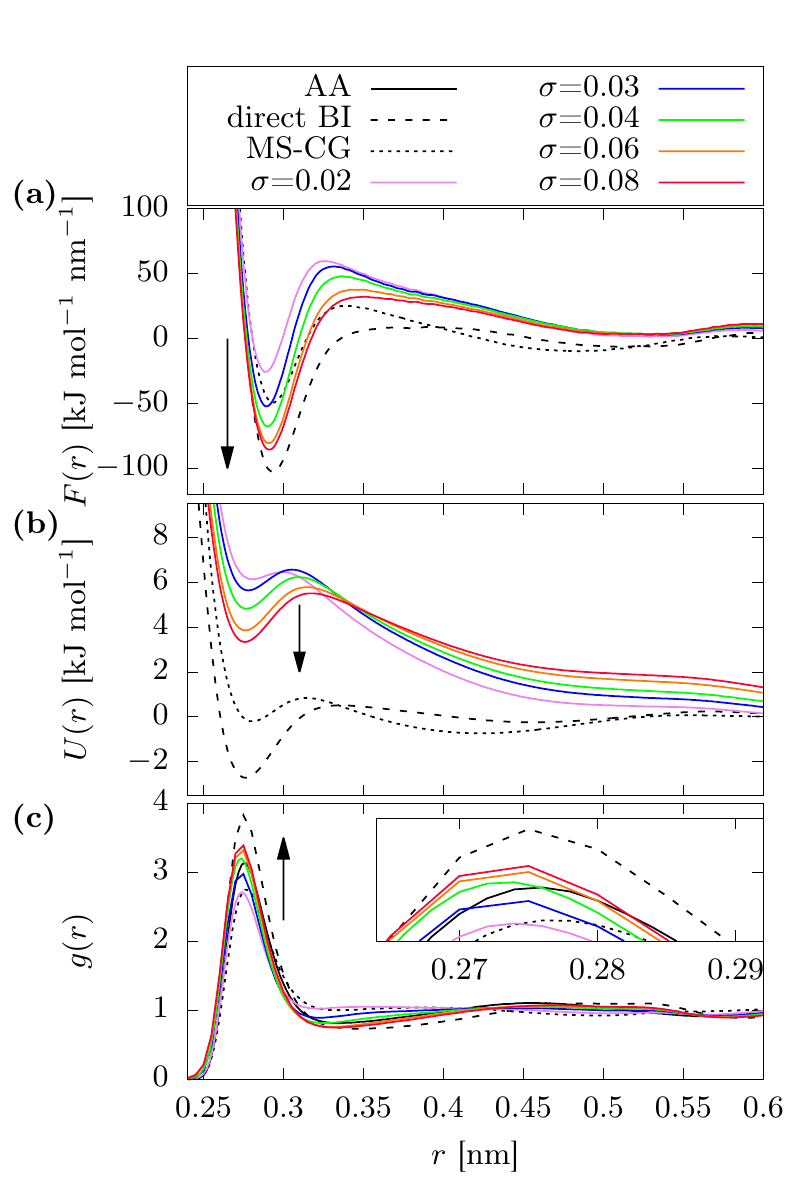}
    \caption{
Forces (a), potentials (b), and RDFs (c) corresponding to the DBI (long dashed curve) and MS-CG (short dashed curve) models as well as several models obtained from smoothing the AA cross-correlation matrix using a Gaussian filter with various standard deviations, ranging from 0.02 to 0.08~nm (solid colored curves).
The AA RDF is presented as the solid black curve in panel (c).
Note that $\sigma \rightarrow \infty$ coincides with DBI.}
\label{fig:gauss}
\end{figure}

Panels (a) and (b) of Figure~\ref{fig:gauss} present the forces and corresponding potentials, respectively, obtained using the filtered matrices with varying standard deviations (solid colored curves).
In comparison with the DBI (long dashed black curve) and MS-CG (short dashed black curve) interactions, the forces and potentials are shifted upward (i.e., more repulsive overall).
This shift appears to be due to the spread of correlations into the hard core and might be counteracted by employing the filtering scheme with alternative boundary conditions.
The forces also demonstrate a systematic transition to larger attractions within the first solvation shell as the amount of smoothing is increased (i.e., larger standard deviations).
This leads to a corresponding increase in the magnitude of the first solvation shell peak of the RDF (Figure~\ref{fig:gauss}(c)).
The model obtained from the smoothed matrix with a standard deviation of $\sigma$ = 0.04~nm results in the most accurate description of the AA RDF.
In contrast to the models obtained with the iterative inversion method, the RDFs generated by the Gaussian filter models exhibit a second solvation shell peak more closely resembling that of the AA RDF.

\begin{figure}[ht!]
    \includegraphics[width=8.5cm]{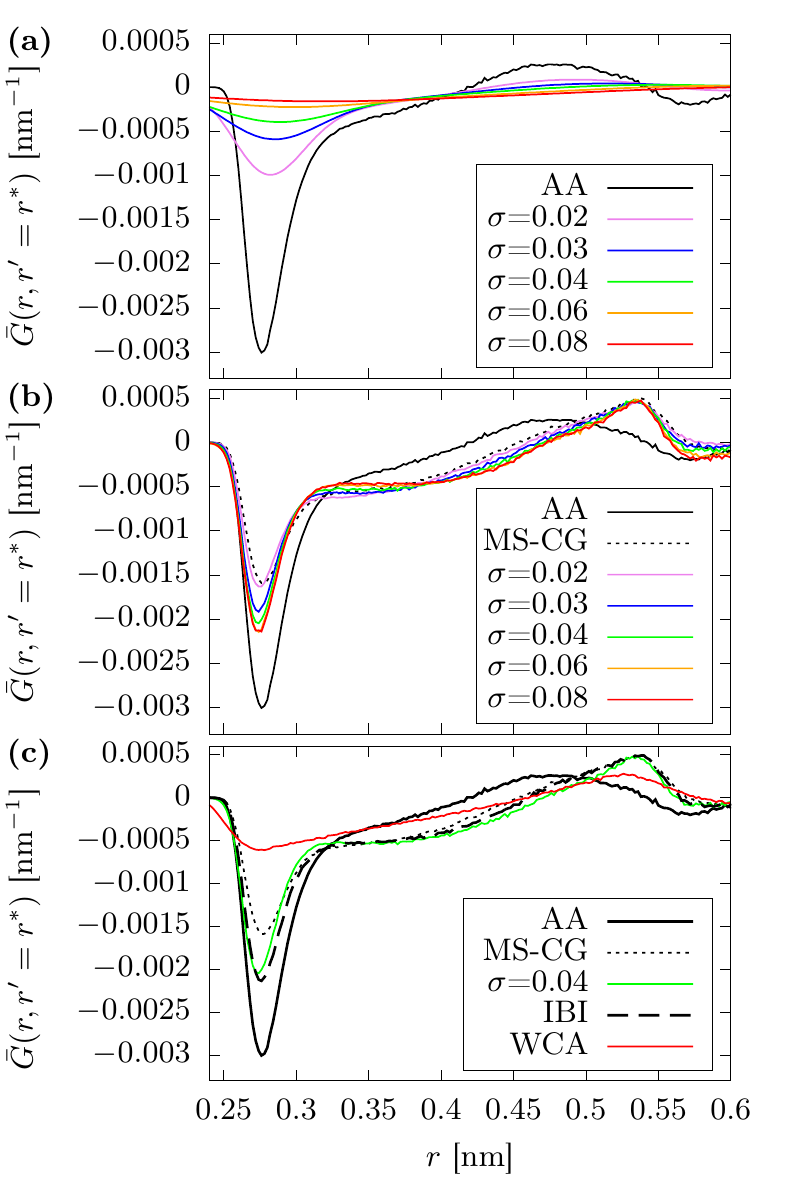}
      \caption{
Cross-sections of the cross-correlation matrix at $r' = r^\ast$ = 0.276~nm (as indicated by the cyan line in Figure~\ref{fig:overview}(b)).
(a) ''Predicted correlations'' -- Cross correlations originally generated by the AA model (solid black curve) and transformed using the Gaussian filtering with various standard deviations (solid colored curves).
These correlations were used to determine the MS-CG and Gaussian filtered models, respectively.
(b) and (c) ``Simulated correlations'' -- Cross correlations generated by the AA (solid black curve), Gaussian filtered (solid colored curves in panel (b)), MS-CG (short dashed black curve), IBI (long dashed black curve), and WCA (solid red curve in panel (c)) models.
The WCA model corresponds to a purely repulsive interaction that was constructed to reproduce the correct solvation structure placement.
}
        \label{fig:matrixxsection}
    \end{figure}

Figure \ref{fig:matrixxsection}(a) presents cross-sections of the correlation matrices (as indicated by the cyan line in panel (b) of Figure~\ref{fig:overview}) obtained by filtering the AA matrix with varying standard deviations (solid colored curves).
The filter reduces the magnitude of the correlation extrema, while retaining their placement.
For standard deviations greater than $\sigma$ = 0.04~nm, only the first solvation shell feature remains, while the description of the second solvation shell correlations has been eliminated.
In the case that the cross-correlation matrix used to solve Equation~\ref{eq:setlineq} truly resembles correlations that can be generated by the CG model, the g-YBG equations become self consistent---the resulting force field will generate the same set of correlations when simulated.
To assess the extent of self-consistency, Figure~\ref{fig:matrixxsection}(b) presents the correlation cross-sections generated from simulations of the AA (solid black curve), MS-CG (short dashed black curve), and Gaussian filtered (solid colored curves) models.
Comparison of the ``predicted'' correlations (panel (a)) and simulated correlations (panel (b)) characterize the extent of self-consistency.
The green curves in panels (a) and (b) demonstrate a considerable lack of self-consistency in the g-YBG equations that determined the $\sigma$ = 0.04~nm model, which most closely reproduces the AA RDF.
Panel (b) also highlights a certain amount of insensitivity of the CG cross correlations with respect to the direct contribution from the CG interactions (i.e., panel (a) of Figure~\ref{fig:gauss}), although the magnitude of the correlation extremum within the first solvation shell does increase somewhat with increasing standard deviation for filtering (i.e., with increasing attractive forces within the first solvation shell).
Therefore the differences in the RDFs generated by the MS-CG and the various Gaussian filtered models appear to be largely due to differences in the direct force, while the indirect contributions from the liquid environment seem to remain approximately fixed.

Panel (b) of Figure \ref{fig:matrixxsection} also demonstrates significant differences in the CG correlations relative to the AA reference.
In particular, the CG models exhibit a correlation extremum within the first solvation shell ($r\approx$ 0.276~nm) with about $2/3$ the magnitude of the AA feature.
Even more noticeable is the qualitative difference in the CG correlation extrema within the second solvation shell ($r\approx$ 0.48~nm) relative to the AA correlations.
This peak is shifted to larger distances for the CG models and is also sharper than in the AA model, supposedly due to the isotropic nature of the CG sites.
This claim is verified in Figure~\ref{fig:matrixxsection}(c), where we present the correlations generated by a purely repulsive, Weeks-Chandler-Andersen (WCA) potential constructed with the correct placement of the solvation structure (solid red curve), which displays correlation extrema with similar placement, albeit with reduced magnitude.
Conversely, the placement and shape of the second correlation extremum generated by the AA model can be attributed to the presence of interactions, e.g., hydrogen bonding, which are anisotropic at the CG level of resolution.
Due to the isotropic nature of a coarse-grained site, it is impossible to reproduce the tetrahedral ordering and, consequently, the second solvation correlation feature demonstrated by the AA model, while retaining the correct radial distribution function, as previously demonstrated by Wang et al~\cite{Wang:2009ol}. 

Figure~\ref{fig:matrixxsection}(c) also presents the correlation cross-section generated by the IBI model (long dashed black curve).
These correlations represent the correlations necessary for a self-consistent g-YBG equation with the AA structural correlations $\bb^{\rm AA}$---or equivalently the AA RDF---as the target structure. 
The correlation cross-sections generated by the IBI and $\sigma$ = 0.04~nm models, as well as those generated by the $\sigma$ = 0.02~nm and MS-CG models (panel (b) of Figure~\ref{fig:matrixxsection}), are quite similar and yield similar RDFs (panel (c) of Figure~\ref{fig:gauss}), despite having markedly different force fields.
To investigate this apparent inconsistency, we analyzed various contributions to the mean force (Equation~\ref{eq:ppmf}) for each model, with respect to the contributions in the IBI model.
In particular, we first decomposed $\bb$ as
\begin{eqnarray}
\bb^{(i)} &=& \bG^{(i)} \bphi^{(i)} \nonumber \\ 
          &=& \bar{\bm g}^{(i)} \circ \bphi^{(i)} + \bar{\bG}^{(i)} \bphi^{(i)} \nonumber \\ 
          &=& \bar{\bm g}^{(i)} \circ \bphi^{(i)} + (\bar{\bG}^{\rm IBI}+\delta \bar{\bG}^{(i)} )(\bphi^{\rm IBI}+\delta \bphi^{(i)}) \nonumber \\
&=& \bar{\bm g}^{(i)} \circ \bphi^{(i)} + \bar{\bG}^{\rm IBI}\bphi^{\rm IBI} + \bar{\bG}^{\rm IBI}\delta \bphi^{(i)} \nonumber \\
& & + \delta \bar{\bG}^{(i)} \bphi^{(i)} \,\, ,
\label{eq:b-decomp}
\end{eqnarray}
where $\circ$ indicates elementwise vector multiplication.
Dividing Equation~\ref{eq:b-decomp} by $\bar{\bm g}$ and using Equation~\ref{eq:ppmf} along with some algebraic rearrangements (see Appendix~\ref{appA} for details), 
the difference between the mean forces generated by model $i$ and the IBI model, $\delta \left ( -\bw'^{(i)} \right ) = \left ( -\bw'^{(i)} \right ) - \left ( -\bw'^{\rm IBI} \right )$, can be written:
\begin{eqnarray}
\delta (- \bw'^{(i)}) &=& \delta \bphi^{(i)} + \left(\frac{\bar{\bm g}^{\rm IBI}}{\bar{\bm g}^{(i)}} - 1 \right) \frac{1}{\bar{\bm g}^{\rm IBI}} \bar{\bG}^{\rm IBI}\bphi^{\rm IBI} \nonumber \\
& & + \frac{1}{\bar{\bm g}^{(i)}} \bar{\bG}^{\rm IBI}\delta \bphi^{(i)} + \frac{1}{\bar{\bm g}^{(i)}} \delta \bar{\bG}^{(i)} \bphi^{(i)} \,\, .
\label{eq:dw-decomp}
\end{eqnarray}
The first term in Equation~\ref{eq:dw-decomp} quantifies the direct contribution to differences in the mean force between the models, while the remaining three terms quantify indirect contributions.
The second term arises due to differences in the RDF between the models, vanishing in the limit of identical pair structure. 
The third and fourth terms (abbreviated $\bar{\bG}^{\rm IBI} \delta \bphi^{(i)}$ and $\delta \bar{\bG}^{(i)} \bphi^{\rm IBI}$, respectively) quantify differences in the mean force due to changing the direct and indirect contributions to $\bb$, respectively.
The second and fourth terms were found to display mirroring behavior in all cases as well as opposing divergence at very short distances.
For this reason, we have combined these two terms in the following analysis, referred to by simply $\delta \bar{\bG}^{(i)} \bphi^{\rm IBI}$ for convenience.

\begin{figure}[ht!]
\includegraphics[width=8.5cm]{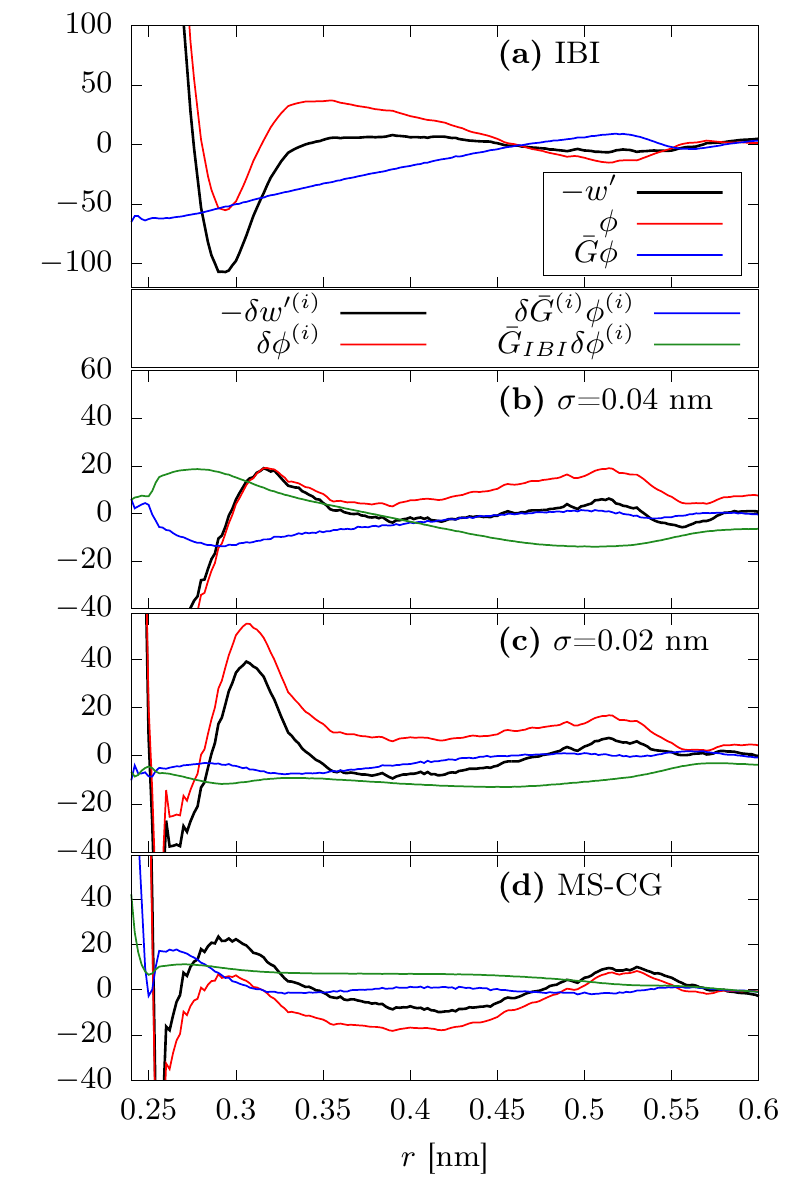}
\caption{
(a) The pair mean force, $-w'(r)$ generated by the IBI model is decomposed into direct ($\phi$) and indirect ($\bar{G}\phi$) contributions.
The difference between the pair mean force generated by the (b) $\sigma$ = 0.04~nm, (c) $\sigma$ = 0.02~nm, and (d) MS-CG models, relative to the IBI model, is decomposed into four contributions (Equation~\ref{eq:dw-decomp}): 
(i) the difference between the direct forces ($\delta \phi^{(i)}$),
(ii) the indirect contribution from differences in the resulting RDFs ($\delta g^{(i)} \bar{\bG}^{\rm IBI} \bphi^{\rm IBI}$),
(iii) the indirect contribution from changes in the direct force relative to IBI ($\bar{G}^{\rm IBI} \delta \phi^{(i)}$),
(iv) the indirect contribution from changes in the cross correlations relative to IBI ($\delta \bar{G}^{(i)} \phi^{\rm IBI}$).
The contributions from (ii) and (iv) are presented as a single net contribution, labeled $\delta \bar{G}^{(i)} \phi^{\rm IBI}$.
Note that panels (b)-(d) are on the same scale, and the tick label at 60 on the y-axis was removed in panels (c) and (d) for clarity.
All y-axes are in units of kJ mol$^{-1}$ nm$^{-1}$.}
\label{fig:dw}
\end{figure}

Figure~\ref{fig:dw} presents decompositions of the pair mean force, $-w'(r)$.
Panel (a) shows the direct (red curve) and indirect (blue curve) contributions to the pair mean force (black curve) for the IBI model, while panels (b)-(d) present the four contributions (Equation~\ref{eq:dw-decomp}) to the difference in $-w'(r)$ between the $\sigma$ = 0.04~nm, $\sigma$ = 0.02~nm and MS-CG models, respectively, relative to the IBI model.
The $\sigma$ = 0.04~nm model (panel (b)) illustrates a small change in the pair mean force relative to the IBI model (black curve), as already shown in Figure~\ref{fig:gauss}(c) in terms of differences in the RDF.
Note that the largest differences occur adjacent to the location of the first solvation shell peak in the RDF ($r \approx$ 0.276~nm), while the difference at $r \approx$ 0.276~nm is close to zero.
The differences in the mean force at short distances are almost entirely due to the difference in the direct force (red curve).
Contrastingly, at larger distances there is almost no change in the mean force, even though the change in the direct force is significant.
The change in the direct force is compensated by the indirect contributions due to these changes in the direct force ($\bG^{\rm IBI} \delta \bphi^{(i)}$, green curve).
That is, the nature of the cross correlations automatically suppress these particular changes to the direct force for distances larger than $r \approx$ 0.4~nm.
Additionally, the indirect contributions due to changes in the cross correlations ($\delta \bG^{(i)} \bphi^{(i)}$, blue curve) appear to be negligible at larger distances, as expected from the similarity of the correlation cross-sections between the $\sigma$ = 0.04~nm and IBI models presented in Figure~\ref{fig:matrixxsection}(c).
These contributions do become significant at short distances but are counteracted by $\bG^{\rm IBI} \delta \bphi^{(i)}$.

Panels (c) and (d) show more pronounced changes in the pair mean forces relative to the IBI model, as expected from the $\sigma$ = 0.02~nm and MS-CG RDFs presented in Figure~\ref{fig:gauss}(c).
Note that while the MS-CG model demonstrates a similar magnitude of difference in the mean force for short distances as the $\sigma$ = 0.04~nm model, the placement of this difference corresponds to the first solvation shell peak of the AA RDF.
Moreover, both the $\sigma$ = 0.02~nm and MS-CG models display more significant differences at distances larger than $r \approx$ 0.35~nm.
The large changes in the $\sigma$ = 0.02~nm direct force, relative to the IBI model, are partially compensated for by the $\bG^{\rm IBI} \delta \bphi^{(i)}$ contributions throughout the entire distance range.
The $\delta \bG^{(i)} \bphi^{(i)}$ contributions are very small, as expected from the difference in the correlation cross-sections in Figure~\ref{fig:matrixxsection}(b), but also compensate for changes in the direct force at short distances.

The MS-CG model demonstrates somewhat different behavior.
The change in the direct force at short distances ($r < 0.325$~nm) is relatively small. 
However, the direct force is more distinct from the IBI force at intermediate distances ($0.325$~nm $< r < 0.475$~nm) and, perhaps more importantly, demonstrates a parabolic change in this region.
In contrast, the changes in the direct force for the $\sigma$ = 0.02~nm and $\sigma$ = 0.04~nm models in this region are quite flat.
While the $\bG^{\rm IBI} \delta \bphi^{(i)}$ contributions somewhat suppress the changes in the direct force within the intermediate region, they exacerbate the changes at short ($r \approx$ 0.3~nm) and long ($r \approx$ 0.525~nm) distances.
The $\delta \bG^{(i)} \bphi^{(i)}$ contributions are again small, but contribute to changes in the mean force at short distances.

This analysis indicates that the structure of the liquid can be very insensitive to particular changes in the direct force and rather sensitive to other changes.
Changes in the direct force at larger distances appear to be typically suppressed, in conjunction with Weeks-Chandler-Andersen theory, due to the corresponding change in indirect contributions from the environment.
However, depending on the nature of the changes in the direct force at larger distances, a significant change in the mean force can persist.
To investigate this in more detail, we considered a set of distinct models which all approximately reproduced the AA RDF. 
To construct these models, we used the difference between the $\sigma$ = 0.04~nm and the IBI potentials, $\delta U^{\sigma = 0.04} = U^{\sigma = 0.04} - U^{\rm IBI}$, as a proxy for structurally-invariant changes in the interaction.
One could also take a more rigorous approach by examining the eigenspectrum of the cross-correlation matrix~\cite{Rudzinski:2012vn}.
We then determined three new models as follows:
\begin{enumerate}
\item[(i)] $U^{\rm above} = U^{\sigma = 0.04} + 0.5\delta U^{\sigma = 0.04}$ \,\, ,
\item[(ii)] $U^{\rm intermediate} = U^{\sigma = 0.04} - 0.5\delta U^{\sigma = 0.04}$ \,\, ,
\item[(iii)] $U^{\rm below} = U^{\rm IBI} - 0.5\delta U^{\sigma = 0.04}$ \,\, .
\end{enumerate}

\begin{figure}[ht!]
\includegraphics[width=8.5cm]{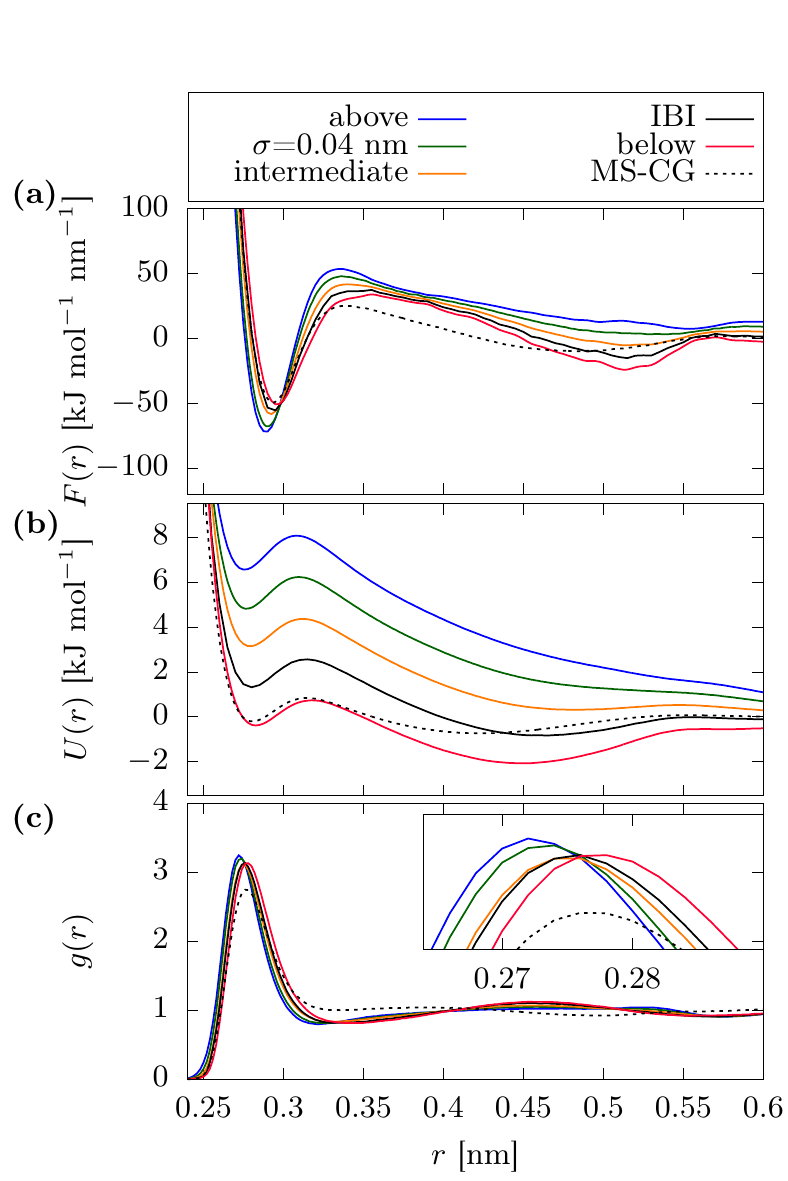}
\caption{
Forces (a), potentials (b), and RDFs (c) corresponding to various models constructed from the difference between the $\sigma = 0.04$ and IBI forces (see main text for details). The RDF of the IBI model coincides with the AA RDF.}
\label{fig:constr}
\end{figure}

Panels (a) and (b) of Figure~\ref{fig:constr} present the forces and corresponding potentials, respectively, for these three models as well as the $\sigma = 0.04$, IBI, and MS-CG models.
Figure~\ref{fig:constr}(c) demonstrates that the pair structure is minimally perturbed by these rather large changes in the potentials.
Conversely, the MS-CG and ``below'' models have very similar interaction potentials at short distances (panel (b)) but generate significantly different RDFs (panel (c)), indicating that a balance between the first and second solvation features in the interactions lead to the correct pair structure for the ``structurally-invariant'' models.
This result is reminiscent of previous work from Yan \etal~\cite{Yan:2006,Yan:2008} (expanded upon by Chaimovich and Shell~\cite{Chaimovich:2009tw}), who demonstrated that the ratio of two characteristic length scales determine water-like behavior for core-softened potentials by enforcing the proper migration of water molecules from the second to the first solvation shell.

\begin{figure}[ht!]
\includegraphics[width=8.5cm]{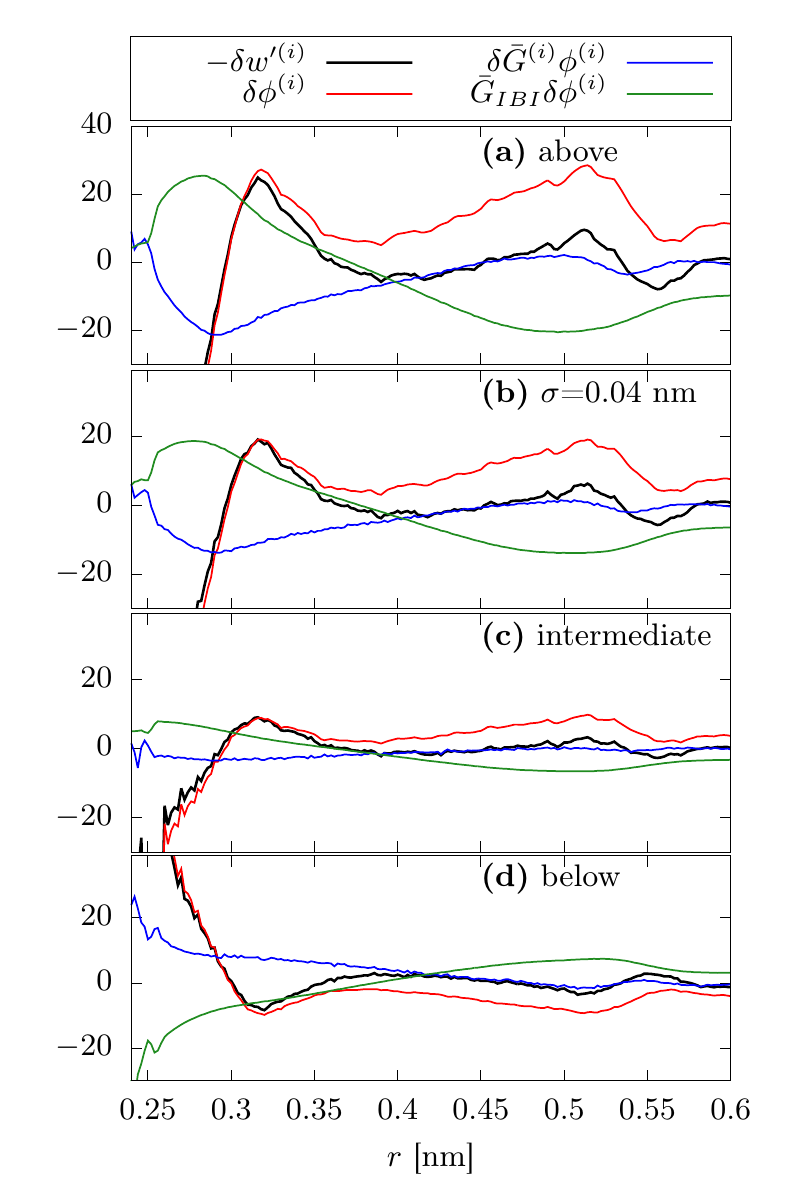}
\caption{
The difference between the pair mean force generated by the (a) ``above'', (b) $\sigma$ = 0.04~nm, (c) ``intermediate'', and (d) ``below'' models, relative to the IBI model, is decomposed into three contributions (Equation~\ref{eq:dw-decomp}):
(i) the difference between the direct forces ($\delta \phi^{(i)}$),
(ii) the indirect contribution from differences in the resulting RDFs ($\delta g^{(i)} \bar{\bG}^{\rm IBI} \bphi^{\rm IBI}$),
(iii) the indirect contribution from changes in the direct force relative to IBI ($\bar{G}^{\rm IBI} \delta \phi^{(i)}$),
(iv) the indirect contribution from changes in the cross correlations relative to IBI ($\delta \bar{G}^{(i)} \phi^{\rm IBI}$).
The contributions from (ii) and (iv) are presented as a single net contribution, labeled $\delta \bar{G}^{(i)} \phi^{\rm IBI}$.
Note that panels (b)-(d) are on the same scale, and the tick label at 60 on the y-axis was removed in panels (c) and (d) for clarity.
All y-axes are in units of kJ mol$^{-1}$ nm$^{-1}$.}
\label{fig:constr_dw}
\end{figure}

Figure~\ref{fig:constr_dw} presents decompositions of the difference in the pair mean force for these models, relative to the IBI model.
As in Figure~\ref{fig:dw}, three contributions are considered.
The $\sigma$ = 0.04~nm model, presented in panel (b), was already analyzed above (Figure~\ref{fig:dw}(b)).
Panels (a), (c), and (d) show analogous results.
First, notice that the constructed change in the potentials results in a relatively small change in the direct force at short distances (Figure~\ref{fig:constr}(a)).
The red curves in Figure~\ref{fig:constr_dw} demonstrate that the short-distance changes that do persist result in a corresponding change in the mean forces (black curves).
On the other hand, the direct forces demonstrate rather significant differences for larger distances.
However, these differences are not passed on to the mean force, due to a cancellation with the indirect contributions from these change in the direct force ($\bG^{\rm IBI} \delta \bphi^{(i)}$, green curves).
This cancellation clearly does not occur for the MS-CG model, as the difference in the forces, relative to the ``below'' model, lead to significant deviations in the mean force (and thus also the RDF).
As above, the indirect contributions due to changes in the cross correlations ($\delta \bG^{(i)} \bphi^{(i)}$, blue curves) are small at large distances and cancel with the $\bG^{\rm IBI} \delta \bphi^{(i)}$ term at short distances, resulting in minimal impact on the resulting mean force.

These results have serious consequences for methods which use cross-correlation information to inform the optimal interaction parameters for CG models.
In particular, it seems that the overall description of the correlations---feature position and relative feature magnitude---may be more important for accurately determining the optimal set of parameters than finer features of the correlations.
A feature that is misplaced, i.e., whose extremum position cannot be reproduced by the CG model, may be detrimental to the quality of the resulting model, even if other correlation features are accurate.
Indeed, this is exactly the case for the second solvation shell feature generated by the AA model of water.
This extends previous results from Rudzinski and Noid~\cite{Rudzinski:2014} who showed that in cases where the AA cross correlations cannot be reproduced by the CG model, an artificial cross-correlation matrix---constructed by assuming statistical independence of the interactions---can recover a model which reproduces the 1-D distributions along degrees of freedom that govern the CG interactions, despite not corresponding to a self-consistent set of g-YBG equations.

From the results thus far, the precise role that specific differences in the cross correlations (with respect to the AA correlations) play in determining structural deficiencies of the model remains unclear.
To investigate the role of the differences within the first and second solvation shell regions independently, we construct three artificial correlation matrices: 
\begin{compactenum}[(i)]
\item\label{it:1} To determine the impact of the placement and shape of the second solvation shell correlations, we construct a matrix by combining AA correlations for distances smaller than 0.32~nm with IBI correlations for all larger distances (AA+IBI).
\item\label{it:2} To determine the impact of the magnitude of the first solvation shell correlations, we construct a matrix by taking the AA correlations for distances smaller than 0.32~nm while setting all correlations at larger distances to zero (AA+0).
\item\label{it:3} To verify the effect of the magnitude of the first solvation shell, we construct a matrix by taking the IBI correlations for distances smaller than 0.32~nm while setting all correlations at larger distances to zero (IBI+0).
\end{compactenum}

\begin{figure}[ht!]
\includegraphics[width=8.5cm]{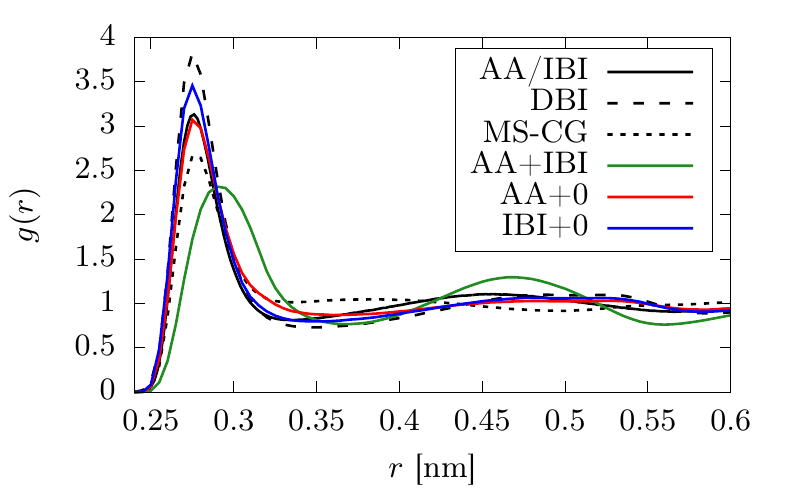}
\caption{
RDFs generated by the AA, DBI, and MS-CG models compared with three models constructed using artificial cross-correlation matrices (see main text for details).}
\label{fig:rdf}
\end{figure}

Figure~\ref{fig:rdf} presents the RDFs generated by models constructed from the correlation matrices described in (i)-(iii).
The RDF generated by the AA+IBI model (solid green curve) demonstrates drastic differences from the AA RDF (solid black curve).
Both the position and magnitude of the first solvation shell peak show significant deviations, while the second solvation shell peak is at the correct distance but overstructured.
This demonstrates that, given the proper correlations at larger distances, the magnitude of the first correlation extremum is essential for even qualitative reproduction of the RDF.
Although it is possible that these large differences in the RDF generated by the AA+IBI could be due to numerical artifacts from combining two correlation matrices, we think that this is unlikely, since there appears to be a smooth transition between correlations in the interface region.
Moreover, the matrices used to determine the AA+0 and IBI+0 models contain a discrete jump at the transition region, but lead to much smaller deviations in the RDFs, as described further below.

Surprisingly, the RDF generated by the AA+0 model (solid red curve) reproduces the AA RDF quite accurately.
In the absence of second correlation feature, the increased magnitude of the AA first correlation extremum relative to the IBI correlations apparently provides the appropriate effective correlations to determine a model which reproduces the AA RDF.
This is further verified by the IBI+0 model (solid blue curve) which introduces discrepancies into the RDF.
These tests provide insight into the results from the Gaussian filtered models: The $\sigma =$~0.4~nm filtering produces correlations that give rise to the appropriate interaction potential by eliminating the AA second solvation correlation features, which cannot be reproduced by the CG model, while also providing a first solvation correlation which, despite its drastic difference from the ``true'' CG correlations, provides the appropriate effective correlations. 

\section{Summary and Conclusions}
\label{sec:conc}

The generalized Yvon-Born-Green (g-YBG) framework, in conjunction with the multiscale coarse-graining (MS-CG) method, describes how cross correlations between the degrees of freedom governing interactions in the coarse-grained (CG) potential inform the set of interaction parameters that give rise to the correct 1-D distributions along these degrees of freedom with respect to an AA reference model.
Although this procedure is exact in the limit of a complete basis set (i.e., infinitely complex interactions), the molecular mechanics potentials typically employed to represent CG interactions are highly deficient for describing AA structural correlations.
The MS-CG method employs the AA cross correlations as a proxy for CG correlations, which can result in severe errors in the resulting model, but can also identify deficiencies in the chosen CG representation and set of interactions.
If one is only interested in reproducing the set of 1-D structural correlation functions along the CG degrees of freedom, iterative procedures can be applied to determine the appropriate set of parameters.
However, the convergence properties of these methods may be problematic as the complexity of the system increases.

The present work has proposed two methods for transforming AA cross correlations to more accurately reflect the correlations that can be reproduced with a given CG representation and set of interactions.
These approaches provide a direct route for determination of the set of CG interaction parameters that will give rise to the appropriate set of 1-D structural correlation functions, even when the AA cross correlations are too complex for the CG model to exactly reproduce.
The proposed iterative inversion method leverages the natural structure of the cross-correlation matrix to prioritize correlation features, through a truncated Neumann series expansion to solve the g-YBG equations.
While this provides a rigorous interpolation between the MS-CG and direct Boltzmann inversion models, our results demonstrate that this method is highly limited when there exist dominant correlation features which cannot be reproduced by the CG model.
As such, this approach may be more useful in cases where several, smaller discrepancies in the cross correlations result in significant errors in the resulting CG structural distributions.
The proposed Gaussian filtering method is also capable of softening the features of the AA cross-correlation matrix in cases where the AA and CG correlations are qualitatively comparable.
However, this method is also capable of eradicating correlations in cases where the AA and CG models generate qualitatively different features, e.g., the second solvation shell feature for the one-site water model considered in this work.

Analysis of various contributions to the pair mean force indicates that differences between models are predominantly due to differences in the direct forces between pairs of CG sites, while the cross correlations (between triplets of sites) are somewhat invariant.
Moreover, the differences that are observed between cross correlations have little impact on the resulting mean force, indicating the precise form of the cross correlations may not be crucial for informing the CG parametrization, as long as the dominant features of the correlations are represented.
Our results also demonstrate that the cross correlations seem to generally suppress changes in the direct force at larger distances, consistent with Weeks-Chandler-Anderson theory.
However, depending on the precise nature of the change, adjustments in the force at larger distances can also significantly impact the RDF.
Overall, there is a balance between correlation features (e.g., between features in the first and second solvation shell in this case) which is essential for determining interaction parameters that give rise to the proper structural correlation functions.

%These results motivate the adoption of simple transformation schemes to adjust the AA correlations to more accurately represent CG correlations, within the context of the MS-CG/g-YBG method.
%In particular,
In cases where the MS-CG model fails to accurately reproduce the 1-D distribution functions and the iterative techniques present convergence problems, methods to systematically simplify the AA cross correlations provide an alternative route for constructing structurally-accurate models.
It is easy to imagine extending the methods proposed in this work, e.g., by introducing non-uniform spacing or alternative functional forms for filtering.
These approaches may be especially useful for multi-component systems with many different interactions or the refinement of existing models where portions of the force field should remain fixed.
In these situations, relatively small differences in individual cross correlations can accumulate and propagate into large errors in the resulting model.
Furthermore, for CG models that incorporate higher-order interactions, the determination of cross correlations may become prohibitively expensive, while iterative refinement may also pose computational challenges.
In this case, the generation of ``artificial'' correlations using the intuition provided by the present results may provide an alternative route for efficient and accurate parametrizations.

\section{Acknowledgments}
The authors thank Christoph Scherer and Denis Andrienko for supplying the atomistic simulations of water along with the corresponding IBI potential.
The authors are also grateful to Robin Cortes-Huerto and Christoph Scherer for critical reading of the manuscript.
SJW thanks Christoph Scherer for insightful discussions.
JFR thanks Will Noid for the introduction to the Neumann series expansion and for many insightful discussions on bottom-up coarse-graining.
SJW was funded by the TRR 146 Collaborative Research Center of the Deutsche Forschungsgemeinschaft (DFG).
TB was funded by the Emmy Noether program.
JFR was partially funded by a postdoctoral fellowship from the Alexander von Humboldt foundation.

\appendix

\section{Decomposition of the mean force}
\label{appA}

Consider two models $\bphi^{(i)}$ and $\bphi^{(j)}$ with (exact) g-YBG equations: $\bb^{(i)} = \bG^{(i)} \bphi^{(i)}$ and $\bb^{(j)} = \bG^{(j)} \bphi^{(j)}$.
The force/structural correlation function of model $i$, $\bb^{(i)}$, can be decomposed with respect to model $j$ as:
\begin{eqnarray}
\bb^{(i)} &=& \bG^{(i)} \bphi^{(i)} \nonumber \\
          &=& \bar{\bm g}^{(i)} \circ \bphi^{(i)} + \bar{\bG}^{(i)} \bphi^{(i)} \nonumber \\
          &=& \bar{\bm g}^{(i)} \circ \bphi^{(i)} + \left ( \bar{\bG}^{(j)}+\delta \bar{\bG}^{(i)} \right ) \left ( \bphi^{(j)}+\delta \bphi^{(i)} \right ) \nonumber \\
&=& \bar{\bm g}^{(i)} \circ \bphi^{(i)} + \bar{\bG}^{(j)}\bphi^{(j)} + \bar{\bG}^{(j)}\delta \bphi^{(i)} \nonumber \\
& & + \delta \bar{\bG}^{(i)} \bphi^{(j)} + \delta \bar{\bG}^{(i)} \delta \bphi^{(i)} \,\, ,
\label{eq:b-decomp-app}
\end{eqnarray}
where $\circ$ indicates elementwise vector multiplication.
Dividing Equation~\ref{eq:b-decomp-app} by $\bar{\bm g}$ and using Equation~\ref{eq:ppmf}, we can write the pair mean force for model $i$ as:
\begin{eqnarray}
-\bw'^{(i)} &=& \bb^{(i)} / \bar{\bm g}^{(i)} \nonumber \\
           &=& \bphi^{(i)} + \frac{1}{\bar{\bm g}^{(i)}} \bar{\bG}^{(j)}\bphi^{(j)} + \frac{1}{\bar{\bm g}^{(i)}} \bar{\bG}^{(j)}\delta \bphi^{(i)} \nonumber \\
           & & + \frac{1}{\bar{\bm g}^{(i)}} \delta \bar{\bG}^{(i)} \bphi^{(j)} + \frac{1}{\bar{\bm g}^{(i)}}\delta \bar{\bG}^{(i)} \delta \bphi^{(i)}\nonumber \\
           &=& \bphi^{(i)} + \left ( \bphi^{(j)} - \bphi^{(j)} \right ) + \frac{1}{\bar{\bm g}^{(i)}} \bar{\bG}^{(j)}\bphi^{(j)} + \frac{1}{\bar{\bm g}^{(i)}} \bar{\bG}^{(j)}\delta \bphi^{(i)} \nonumber \\
           & & + \frac{1}{\bar{\bm g}^{(i)}} \delta \bar{\bG}^{(i)} \bphi^{(j)} + \frac{1}{\bar{\bm g}^{(i)}} \delta \bar{\bG}^{(i)} \delta \bphi^{(i)}\nonumber \\
           &=& \delta \bphi^{(i)} + \left ( \bphi^{(j)} + \frac{1}{\bar{\bm g}^{(i)}} \bar{\bG}^{(j)}\bphi^{(j)} \right ) + \frac{1}{\bar{\bm g}^{(i)}} \bar{\bG}^{(j)}\delta \bphi^{(i)} \nonumber \\
           & & + \frac{1}{\bar{\bm g}^{(i)}} \delta \bar{\bG}^{(i)} \bphi^{(j)} + \frac{1}{\bar{\bm g}^{(i)}} \delta \bar{\bG}^{(i)} \delta \bphi^{(i)}\nonumber \\
           &=& \delta \bphi^{(i)} + \left ( \bphi^{(j)} + \frac{\bar{\bm g}^{(j)}}{\bar{\bm g}^{(j)}}  \frac{1}{\bar{\bm g}^{(i)}} \bar{\bG}^{(j)}\bphi^{(j)} \right ) + \frac{1}{\bar{\bm g}^{(i)}} \bar{\bG}^{(j)}\delta \bphi^{(i)} \nonumber \\
           & & + \frac{1}{\bar{\bm g}^{(i)}} \delta \bar{\bG}^{(i)} \bphi^{(j)} + \frac{1}{\bar{\bm g}^{(i)}} \delta \bar{\bG}^{(i)} \delta \bphi^{(i)} \nonumber \\
           &=& \delta \bphi^{(i)} \nonumber \\
           & & + \left ( \bphi^{(j)} + \frac{1}{\bar{\bm g}^{(j)}} \bar{\bG}^{(j)}\bphi^{(j)} + \left ( \frac{\bar{\bm g}^{(j)}}{\bar{\bm g}^{(i)}} - 1 \right ) \frac{1}{\bar{\bm g}^{(j)}} \bar{\bG}^{(j)}\bphi^{(j)} \right ) \nonumber \\
           & & + \frac{1}{\bar{\bm g}^{(i)}} \bar{\bG}^{(j)}\delta \bphi^{(i)} + \frac{1}{\bar{\bm g}^{(i)}} \delta \bar{\bG}^{(i)} \bphi^{(j)} + \frac{1}{\bar{\bm g}^{(i)}} \delta \bar{\bG}^{(i)} \delta \bphi^{(i)}\nonumber \\
           &=& \delta \bphi^{(i)} + \left ( -\bw'^{(j)} + \left ( \frac{\bar{\bm g}^{(j)}}{\bar{\bm g}^{(i)}} - 1 \right ) \frac{1}{\bar{\bm g}^{(j)}} \bar{\bG}^{(j)}\bphi^{(j)} \right )  \nonumber \\
           & & + \frac{1}{\bar{\bm g}^{(i)}} \bar{\bG}^{(j)} \delta \bphi^{(i)} + \frac{1}{\bar{\bm g}^{(i)}} \delta \bar{\bG}^{(i)} \bphi^{(j)} \nonumber \\
           & & + \frac{1}{\bar{\bm g}^{(i)}} \delta \bar{\bG}^{(i)} \delta \bphi^{(i)} \,\, .
\label{eq:wp-decomp-app}
\end{eqnarray}
Then, the difference between the mean forces generated by models $i$ and $j$, $\delta \left ( -\bw'^{(i)} \right ) = \left ( -\bw'^{(i)} \right ) - \left ( -\bw'^{(j)} \right )$, can be written:
\begin{eqnarray}
\delta \left ( -\bw'^{(i)} \right ) &=& \delta \bphi^{(i)} + \frac{1}{\bar{\bm g}^{(i)}} \bar{\bG}^{(j)}\delta \bphi^{(i)} + \frac{1}{\bar{\bm g}^{(i)}} \delta \bar{\bG}^{(i)} \bphi^{(j)} \nonumber \\
                      & & + \left ( \frac{\bar{\bm g}^{(j)}}{\bar{\bm g}^{(i)}} - 1 \right ) \frac{1}{\bar{\bm g}^{(j)}} \bar{\bG}^{(j)}\bphi^{(j)} \nonumber \\
                      & & + \frac{1}{\bar{\bm g}^{(i)}} \delta \bar{\bG}^{(i)} \delta \bphi^{(i)} \,\, .
\label{eq:dw-decomp-app}
\end{eqnarray}
For analysis purposes the last three terms were combined when plotting the results.

%=====================================
% References, variant B: external bibliography
%=====================================
\bibliography{references_PSU,references_MPIP}

\begin{thebibliography}{73}%
\makeatletter
\providecommand \@ifxundefined [1]{%
 \@ifx{#1\undefined}
}%
\providecommand \@ifnum [1]{%
 \ifnum #1\expandafter \@firstoftwo
 \else \expandafter \@secondoftwo
 \fi
}%
\providecommand \@ifx [1]{%
 \ifx #1\expandafter \@firstoftwo
 \else \expandafter \@secondoftwo
 \fi
}%
\providecommand \natexlab [1]{#1}%
\providecommand \enquote  [1]{``#1''}%
\providecommand \bibnamefont  [1]{#1}%
\providecommand \bibfnamefont [1]{#1}%
\providecommand \citenamefont [1]{#1}%
\providecommand \href@noop [0]{\@secondoftwo}%
\providecommand \href [0]{\begingroup \@sanitize@url \@href}%
\providecommand \@href[1]{\@@startlink{#1}\@@href}%
\providecommand \@@href[1]{\endgroup#1\@@endlink}%
\providecommand \@sanitize@url [0]{\catcode `\\12\catcode `\$12\catcode
  `\&12\catcode `\#12\catcode `\^12\catcode `\_12\catcode `\%12\relax}%
\providecommand \@@startlink[1]{}%
\providecommand \@@endlink[0]{}%
\providecommand \url  [0]{\begingroup\@sanitize@url \@url }%
\providecommand \@url [1]{\endgroup\@href {#1}{\urlprefix }}%
\providecommand \urlprefix  [0]{URL }%
\providecommand \Eprint [0]{\href }%
\providecommand \doibase [0]{http://dx.doi.org/}%
\providecommand \selectlanguage [0]{\@gobble}%
\providecommand \bibinfo  [0]{\@secondoftwo}%
\providecommand \bibfield  [0]{\@secondoftwo}%
\providecommand \translation [1]{[#1]}%
\providecommand \BibitemOpen [0]{}%
\providecommand \bibitemStop [0]{}%
\providecommand \bibitemNoStop [0]{.\EOS\space}%
\providecommand \EOS [0]{\spacefactor3000\relax}%
\providecommand \BibitemShut  [1]{\csname bibitem#1\endcsname}%
\let\auto@bib@innerbib\@empty
%</preamble>
\bibitem [{\citenamefont {Kirkwood}(1935)}]{Kirkwood:1935ys}%
  \BibitemOpen
  \bibfield  {author} {\bibinfo {author} {\bibfnamefont {J.~G.}\ \bibnamefont
  {Kirkwood}},\ }\href {\doibase {10.1063/1.1749657}} {\bibfield  {journal}
  {\bibinfo  {journal} {J. Chem. Phys.}\ }\textbf {\bibinfo {volume} {3}},\
  \bibinfo {pages} {300} (\bibinfo {year} {1935})}\BibitemShut {NoStop}%
\bibitem [{\citenamefont {Noid}\ \emph
  {et~al.}(2008{\natexlab{a}})\citenamefont {Noid}, \citenamefont {Chu},
  \citenamefont {Ayton}, \citenamefont {Krishna}, \citenamefont {Izvekov},
  \citenamefont {Voth}, \citenamefont {Das},\ and\ \citenamefont
  {Andersen}}]{Noid:2008a}%
  \BibitemOpen
  \bibfield  {author} {\bibinfo {author} {\bibfnamefont {W.~G.}\ \bibnamefont
  {Noid}}, \bibinfo {author} {\bibfnamefont {J.-W.}\ \bibnamefont {Chu}},
  \bibinfo {author} {\bibfnamefont {G.~S.}\ \bibnamefont {Ayton}}, \bibinfo
  {author} {\bibfnamefont {V.}~\bibnamefont {Krishna}}, \bibinfo {author}
  {\bibfnamefont {S.}~\bibnamefont {Izvekov}}, \bibinfo {author} {\bibfnamefont
  {G.~A.}\ \bibnamefont {Voth}}, \bibinfo {author} {\bibfnamefont
  {A.}~\bibnamefont {Das}}, \ and\ \bibinfo {author} {\bibfnamefont {H.~C.}\
  \bibnamefont {Andersen}},\ }\href {\doibase {10.1063/1.2938860}} {\bibfield
  {journal} {\bibinfo  {journal} {J. Chem. Phys.}\ }\textbf {\bibinfo {volume}
  {128}},\ \bibinfo {pages} {244114} (\bibinfo {year}
  {2008}{\natexlab{a}})}\BibitemShut {NoStop}%
\bibitem [{\citenamefont {Izvekov}\ and\ \citenamefont
  {Voth}(2005{\natexlab{a}})}]{Izvekov:2005d}%
  \BibitemOpen
  \bibfield  {author} {\bibinfo {author} {\bibfnamefont {S.}~\bibnamefont
  {Izvekov}}\ and\ \bibinfo {author} {\bibfnamefont {G.~A.}\ \bibnamefont
  {Voth}},\ }\href {\doibase {10.1021/jp044629q}} {\bibfield  {journal}
  {\bibinfo  {journal} {J. Phys. Chem. B}\ }\textbf {\bibinfo {volume} {109}},\
  \bibinfo {pages} {2469} (\bibinfo {year} {2005}{\natexlab{a}})}\BibitemShut
  {NoStop}%
\bibitem [{\citenamefont {Izvekov}\ and\ \citenamefont
  {Voth}(2005{\natexlab{b}})}]{Izvekov:2005e}%
  \BibitemOpen
  \bibfield  {author} {\bibinfo {author} {\bibfnamefont {S.}~\bibnamefont
  {Izvekov}}\ and\ \bibinfo {author} {\bibfnamefont {G.~A.}\ \bibnamefont
  {Voth}},\ }\href {\doibase {10.1063/1.2038787}} {\bibfield  {journal}
  {\bibinfo  {journal} {J. Chem. Phys.}\ }\textbf {\bibinfo {volume} {123}},\
  \bibinfo {pages} {134105} (\bibinfo {year} {2005}{\natexlab{b}})}\BibitemShut
  {NoStop}%
\bibitem [{\citenamefont {Mullinax}\ and\ \citenamefont
  {Noid}(2009)}]{Mullinax:2009b}%
  \BibitemOpen
  \bibfield  {author} {\bibinfo {author} {\bibfnamefont {J.~W.}\ \bibnamefont
  {Mullinax}}\ and\ \bibinfo {author} {\bibfnamefont {W.~G.}\ \bibnamefont
  {Noid}},\ }\href {\doibase {10.1103/PhysRevLett.103.198104}} {\bibfield
  {journal} {\bibinfo  {journal} {Phys. Rev. Lett.}\ }\textbf {\bibinfo
  {volume} {103}},\ \bibinfo {pages} {198104} (\bibinfo {year}
  {2009})}\BibitemShut {NoStop}%
\bibitem [{\citenamefont {Mullinax}\ and\ \citenamefont
  {Noid}(2010)}]{Mullinax:2010}%
  \BibitemOpen
  \bibfield  {author} {\bibinfo {author} {\bibfnamefont {J.~W.}\ \bibnamefont
  {Mullinax}}\ and\ \bibinfo {author} {\bibfnamefont {W.~G.}\ \bibnamefont
  {Noid}},\ }\href {\doibase {10.1021/jp9073976}} {\bibfield  {journal}
  {\bibinfo  {journal} {J. Phys. Chem. C}\ }\textbf {\bibinfo {volume} {114}},\
  \bibinfo {pages} {5661} (\bibinfo {year} {2010})}\BibitemShut {NoStop}%
\bibitem [{\citenamefont {Rudzinski}\ and\ \citenamefont
  {Noid}(2015{\natexlab{a}})}]{Rudzinski:2015}%
  \BibitemOpen
  \bibfield  {author} {\bibinfo {author} {\bibfnamefont {J.~F.}\ \bibnamefont
  {Rudzinski}}\ and\ \bibinfo {author} {\bibfnamefont {W.~G.}\ \bibnamefont
  {Noid}},\ }\href {\doibase {10.1140/epjst/e2015-02408-9}} {\bibfield
  {journal} {\bibinfo  {journal} {Eur. Phys. J. Special Topics}\ }\textbf
  {\bibinfo {volume} {{224}}},\ \bibinfo {pages} {2193} (\bibinfo {year}
  {{2015}}{\natexlab{a}})}\BibitemShut {NoStop}%
\bibitem [{\citenamefont {Ellis}, \citenamefont {Rudzinski},\ and\
  \citenamefont {Noid}(2011)}]{Ellis:2011a}%
  \BibitemOpen
  \bibfield  {author} {\bibinfo {author} {\bibfnamefont {C.~R.}\ \bibnamefont
  {Ellis}}, \bibinfo {author} {\bibfnamefont {J.~F.}\ \bibnamefont
  {Rudzinski}}, \ and\ \bibinfo {author} {\bibfnamefont {W.~G.}\ \bibnamefont
  {Noid}},\ }\href {\doibase {10.1002/mats.201100022}} {\bibfield  {journal}
  {\bibinfo  {journal} {Macromol. Theory Sim.}\ }\textbf {\bibinfo {volume}
  {20}},\ \bibinfo {pages} {478} (\bibinfo {year} {2011})}\BibitemShut
  {NoStop}%
\bibitem [{\citenamefont {Rudzinski}\ and\ \citenamefont
  {Noid}(2012)}]{Rudzinski:2012vn}%
  \BibitemOpen
  \bibfield  {author} {\bibinfo {author} {\bibfnamefont {J.~F.}\ \bibnamefont
  {Rudzinski}}\ and\ \bibinfo {author} {\bibfnamefont {W.~G.}\ \bibnamefont
  {Noid}},\ }\href {\doibase 10.1021/jp3002004} {\bibfield  {journal} {\bibinfo
   {journal} {J. Phys. Chem. B}\ }\textbf {\bibinfo {volume} {116}},\ \bibinfo
  {pages} {8621} (\bibinfo {year} {2012})}\BibitemShut {NoStop}%
\bibitem [{\citenamefont {Rudzinski}\ and\ \citenamefont
  {Noid}(2014)}]{Rudzinski:2014}%
  \BibitemOpen
  \bibfield  {author} {\bibinfo {author} {\bibfnamefont {J.~F.}\ \bibnamefont
  {Rudzinski}}\ and\ \bibinfo {author} {\bibfnamefont {W.~G.}\ \bibnamefont
  {Noid}},\ }\href {\doibase {10.1021/jp501694z}} {\bibfield  {journal}
  {\bibinfo  {journal} {J. Phys. Chem. B}\ }\textbf {\bibinfo {volume} {118}},\
  \bibinfo {pages} {8295} (\bibinfo {year} {2014})}\BibitemShut {NoStop}%
\bibitem [{\citenamefont {Mashayak}, \citenamefont {Miao},\ and\ \citenamefont
  {Aluru}(2018)}]{Mashayak:2018}%
  \BibitemOpen
  \bibfield  {author} {\bibinfo {author} {\bibfnamefont {S.~Y.}\ \bibnamefont
  {Mashayak}}, \bibinfo {author} {\bibfnamefont {L.}~\bibnamefont {Miao}}, \
  and\ \bibinfo {author} {\bibfnamefont {N.~R.}\ \bibnamefont {Aluru}},\
  }\href@noop {} {\bibfield  {journal} {\bibinfo  {journal} {J. Chem. Phys.}\
  }\textbf {\bibinfo {volume} {{148}}},\ \bibinfo {pages} {214105} (\bibinfo
  {year} {{2018}})}\BibitemShut {NoStop}%
\bibitem [{\citenamefont {Noid}(2013{\natexlab{a}})}]{Noid:2013}%
  \BibitemOpen
  \bibfield  {author} {\bibinfo {author} {\bibfnamefont {W.}~\bibnamefont
  {Noid}},\ }\href@noop {} {\bibfield  {journal} {\bibinfo  {journal} {Methods
  Mol. Biol.}\ }\textbf {\bibinfo {volume} {924}},\ \bibinfo {pages} {487}
  (\bibinfo {year} {2013}{\natexlab{a}})}\BibitemShut {NoStop}%
\bibitem [{\citenamefont {Noid}(2013{\natexlab{b}})}]{Noid:2013uq}%
  \BibitemOpen
  \bibfield  {author} {\bibinfo {author} {\bibfnamefont {W.~G.}\ \bibnamefont
  {Noid}},\ }\href {\doibase 10.1063/1.4818908} {\bibfield  {journal} {\bibinfo
   {journal} {J. Chem. Phys.}\ }\textbf {\bibinfo {volume} {139}},\ \bibinfo
  {pages} {090901} (\bibinfo {year} {2013}{\natexlab{b}})}\BibitemShut
  {NoStop}%
\bibitem [{\citenamefont {Tsch{\"o}p}\ \emph {et~al.}(1998)\citenamefont
  {Tsch{\"o}p}, \citenamefont {Kremer}, \citenamefont {Batoulis}, \citenamefont
  {B{\"u}rger},\ and\ \citenamefont {Hahn}}]{Tschop:1998lj}%
  \BibitemOpen
  \bibfield  {author} {\bibinfo {author} {\bibfnamefont {W.}~\bibnamefont
  {Tsch{\"o}p}}, \bibinfo {author} {\bibfnamefont {K.}~\bibnamefont {Kremer}},
  \bibinfo {author} {\bibfnamefont {J.}~\bibnamefont {Batoulis}}, \bibinfo
  {author} {\bibfnamefont {T.}~\bibnamefont {B{\"u}rger}}, \ and\ \bibinfo
  {author} {\bibfnamefont {O.}~\bibnamefont {Hahn}},\ }\href {\doibase
  {10.1002/(SICI)1521-4044(199802)49:2/3<61::AID-APOL61>3.0.CO;2-V}} {\bibfield
   {journal} {\bibinfo  {journal} {Acta Poly.}\ }\textbf {\bibinfo {volume}
  {49}},\ \bibinfo {pages} {61} (\bibinfo {year} {1998})}\BibitemShut {NoStop}%
\bibitem [{\citenamefont {Soper}(1996)}]{Soper:1996ly}%
  \BibitemOpen
  \bibfield  {author} {\bibinfo {author} {\bibfnamefont {A.~K.}\ \bibnamefont
  {Soper}},\ }\href {\doibase {10.1016/0301-0104(95)00357-6}} {\bibfield
  {journal} {\bibinfo  {journal} {Chem. Phys.}\ }\textbf {\bibinfo {volume}
  {202}},\ \bibinfo {pages} {295} (\bibinfo {year} {1996})}\BibitemShut
  {NoStop}%
\bibitem [{\citenamefont {M\"{u}ller-Plathe}(2002)}]{MullerPlathe:2002}%
  \BibitemOpen
  \bibfield  {author} {\bibinfo {author} {\bibfnamefont {F.}~\bibnamefont
  {M\"{u}ller-Plathe}},\ }\href {\doibase
  {10.1002/1439-7641(20020916)3:9<754::AID-CPHC754>3.0.CO;2-U}} {\bibfield
  {journal} {\bibinfo  {journal} {ChemPhysChem}\ }\textbf {\bibinfo {volume}
  {3}},\ \bibinfo {pages} {754} (\bibinfo {year} {2002})}\BibitemShut {NoStop}%
\bibitem [{\citenamefont {Reith}, \citenamefont {Putz},\ and\ \citenamefont
  {M{\"u}ller-Plathe}(2003)}]{Reith:2003tg}%
  \BibitemOpen
  \bibfield  {author} {\bibinfo {author} {\bibfnamefont {D.}~\bibnamefont
  {Reith}}, \bibinfo {author} {\bibfnamefont {M.}~\bibnamefont {Putz}}, \ and\
  \bibinfo {author} {\bibfnamefont {F.}~\bibnamefont {M{\"u}ller-Plathe}},\
  }\href {\doibase 10.1002/jcc.10307} {\bibfield  {journal} {\bibinfo
  {journal} {J. Comp. Chem.}\ }\textbf {\bibinfo {volume} {24}},\ \bibinfo
  {pages} {1624} (\bibinfo {year} {2003})}\BibitemShut {NoStop}%
\bibitem [{\citenamefont {Lyubartsev}\ and\ \citenamefont
  {Laaksonen}(1995)}]{Lyubartsev:1995}%
  \BibitemOpen
  \bibfield  {author} {\bibinfo {author} {\bibfnamefont {A.~P.}\ \bibnamefont
  {Lyubartsev}}\ and\ \bibinfo {author} {\bibfnamefont {A.}~\bibnamefont
  {Laaksonen}},\ }\href {\doibase {10.1103/PhysRevE.52.3730}} {\bibfield
  {journal} {\bibinfo  {journal} {Phys. Rev. E}\ }\textbf {\bibinfo {volume}
  {52}},\ \bibinfo {pages} {3730} (\bibinfo {year} {1995})}\BibitemShut
  {NoStop}%
\bibitem [{\citenamefont {Lyubartsev}\ and\ \citenamefont
  {Laaksonen}(1997)}]{Lyubartsev:1997}%
  \BibitemOpen
  \bibfield  {author} {\bibinfo {author} {\bibfnamefont {A.~P.}\ \bibnamefont
  {Lyubartsev}}\ and\ \bibinfo {author} {\bibfnamefont {A.}~\bibnamefont
  {Laaksonen}},\ }\href {\doibase {10.1103/PhysRevE.55.5689}} {\bibfield
  {journal} {\bibinfo  {journal} {Phys. Rev. E}\ }\textbf {\bibinfo {volume}
  {55}},\ \bibinfo {pages} {5689} (\bibinfo {year} {1997})}\BibitemShut
  {NoStop}%
\bibitem [{\citenamefont {Cho}\ and\ \citenamefont {Chu}(2009)}]{Cho:2009ve}%
  \BibitemOpen
  \bibfield  {author} {\bibinfo {author} {\bibfnamefont {H.~M.}\ \bibnamefont
  {Cho}}\ and\ \bibinfo {author} {\bibfnamefont {J.~W.}\ \bibnamefont {Chu}},\
  }\href {\doibase 10.1063/1.3238547} {\bibfield  {journal} {\bibinfo
  {journal} {J. Chem. Phys.}\ }\textbf {\bibinfo {volume} {131}},\ \bibinfo
  {pages} {134107} (\bibinfo {year} {2009})}\BibitemShut {NoStop}%
\bibitem [{\citenamefont {Lu}, \citenamefont {Dama},\ and\ \citenamefont
  {Voth}(2013)}]{Lu:2013uq}%
  \BibitemOpen
  \bibfield  {author} {\bibinfo {author} {\bibfnamefont {L.}~\bibnamefont
  {Lu}}, \bibinfo {author} {\bibfnamefont {J.~F.}\ \bibnamefont {Dama}}, \ and\
  \bibinfo {author} {\bibfnamefont {G.~A.}\ \bibnamefont {Voth}},\ }\href
  {\doibase 10.1063/1.4811667} {\bibfield  {journal} {\bibinfo  {journal} {J.
  Chem. Phys.}\ }\textbf {\bibinfo {volume} {139}},\ \bibinfo {pages} {121906}
  (\bibinfo {year} {2013})}\BibitemShut {NoStop}%
\bibitem [{\citenamefont {Karimi-Varzaneh}\ and\ \citenamefont
  {M{\"u}ller-Plathe}(2012)}]{Karimi-Varzaneh:2012uq}%
  \BibitemOpen
  \bibfield  {author} {\bibinfo {author} {\bibfnamefont {H.~A.}\ \bibnamefont
  {Karimi-Varzaneh}}\ and\ \bibinfo {author} {\bibfnamefont {F.}~\bibnamefont
  {M{\"u}ller-Plathe}},\ }\href {\doibase 10.1007/128\_2010\_122} {\bibfield
  {journal} {\bibinfo  {journal} {Top. Curr. Chem.}\ }\textbf {\bibinfo
  {volume} {307}},\ \bibinfo {pages} {295} (\bibinfo {year}
  {2012})}\BibitemShut {NoStop}%
\bibitem [{\citenamefont {Qian}\ \emph {et~al.}(2008)\citenamefont {Qian},
  \citenamefont {Carbone}, \citenamefont {Chen}, \citenamefont
  {Karimi-Varzaneh}, \citenamefont {Liew},\ and\ \citenamefont
  {M\"uller-Plathe}}]{Qian:2008}%
  \BibitemOpen
  \bibfield  {author} {\bibinfo {author} {\bibfnamefont {H.-J.}\ \bibnamefont
  {Qian}}, \bibinfo {author} {\bibfnamefont {P.}~\bibnamefont {Carbone}},
  \bibinfo {author} {\bibfnamefont {X.}~\bibnamefont {Chen}}, \bibinfo {author}
  {\bibfnamefont {H.~A.}\ \bibnamefont {Karimi-Varzaneh}}, \bibinfo {author}
  {\bibfnamefont {C.~C.}\ \bibnamefont {Liew}}, \ and\ \bibinfo {author}
  {\bibfnamefont {F.}~\bibnamefont {M\"uller-Plathe}},\ }\href {\doibase
  {10.1021/ma801910r}} {\bibfield  {journal} {\bibinfo  {journal}
  {Macromolecules}\ }\textbf {\bibinfo {volume} {41}},\ \bibinfo {pages} {9919}
  (\bibinfo {year} {2008})}\BibitemShut {NoStop}%
\bibitem [{\citenamefont {Eslami}, \citenamefont {Karimi-Varzaneh},\ and\
  \citenamefont {Mueller-Plathe}(2011)}]{Eslami:2011}%
  \BibitemOpen
  \bibfield  {author} {\bibinfo {author} {\bibfnamefont {H.}~\bibnamefont
  {Eslami}}, \bibinfo {author} {\bibfnamefont {H.~A.}\ \bibnamefont
  {Karimi-Varzaneh}}, \ and\ \bibinfo {author} {\bibfnamefont {F.}~\bibnamefont
  {Mueller-Plathe}},\ }\href@noop {} {\bibfield  {journal} {\bibinfo  {journal}
  {Macromolecules}\ }\textbf {\bibinfo {volume} {{44}}},\ \bibinfo {pages}
  {3117} (\bibinfo {year} {{2011}})}\BibitemShut {NoStop}%
\bibitem [{\citenamefont {Langeloth}\ \emph {et~al.}(2015)\citenamefont
  {Langeloth}, \citenamefont {Sugii}, \citenamefont {Boehm},\ and\
  \citenamefont {Mueller-Plathe}}]{Langeloth:2015}%
  \BibitemOpen
  \bibfield  {author} {\bibinfo {author} {\bibfnamefont {M.}~\bibnamefont
  {Langeloth}}, \bibinfo {author} {\bibfnamefont {T.}~\bibnamefont {Sugii}},
  \bibinfo {author} {\bibfnamefont {M.~C.}\ \bibnamefont {Boehm}}, \ and\
  \bibinfo {author} {\bibfnamefont {F.}~\bibnamefont {Mueller-Plathe}},\
  }\href@noop {} {\bibfield  {journal} {\bibinfo  {journal} {J. Chem. Phys.}\
  }\textbf {\bibinfo {volume} {{143}}},\ \bibinfo {pages} {243158} (\bibinfo
  {year} {{2015}})}\BibitemShut {NoStop}%
\bibitem [{\citenamefont {Peter}, \citenamefont {Delle~Site},\ and\
  \citenamefont {Kremer}(2008)}]{Peter:2008vc}%
  \BibitemOpen
  \bibfield  {author} {\bibinfo {author} {\bibfnamefont {C.}~\bibnamefont
  {Peter}}, \bibinfo {author} {\bibfnamefont {L.}~\bibnamefont {Delle~Site}}, \
  and\ \bibinfo {author} {\bibfnamefont {K.}~\bibnamefont {Kremer}},\ }\href
  {\doibase 10.1039/b717324e} {\bibfield  {journal} {\bibinfo  {journal} {Soft
  Matter}\ }\textbf {\bibinfo {volume} {4}},\ \bibinfo {pages} {859} (\bibinfo
  {year} {2008})}\BibitemShut {NoStop}%
\bibitem [{\citenamefont {Savelyev}\ and\ \citenamefont
  {Papoian}(2009{\natexlab{a}})}]{Savelyev:2009a}%
  \BibitemOpen
  \bibfield  {author} {\bibinfo {author} {\bibfnamefont {A.}~\bibnamefont
  {Savelyev}}\ and\ \bibinfo {author} {\bibfnamefont {G.~A.}\ \bibnamefont
  {Papoian}},\ }\href {\doibase {10.1016/j.bpj.2009.02.067}} {\bibfield
  {journal} {\bibinfo  {journal} {Biophys. J.}\ }\textbf {\bibinfo {volume}
  {96}},\ \bibinfo {pages} {4044} (\bibinfo {year}
  {2009}{\natexlab{a}})}\BibitemShut {NoStop}%
\bibitem [{\citenamefont {Savelyev}\ and\ \citenamefont
  {Papoian}(2010)}]{Savelyev:2010}%
  \BibitemOpen
  \bibfield  {author} {\bibinfo {author} {\bibfnamefont {A.}~\bibnamefont
  {Savelyev}}\ and\ \bibinfo {author} {\bibfnamefont {G.~A.}\ \bibnamefont
  {Papoian}},\ }\href {\doibase {10.1073/pnas.1001163107}} {\bibfield
  {journal} {\bibinfo  {journal} {Proc. Natl. Acad. Sci. USA}\ }\textbf
  {\bibinfo {volume} {107}},\ \bibinfo {pages} {20340} (\bibinfo {year}
  {2010})}\BibitemShut {NoStop}%
\bibitem [{\citenamefont {Savelyev}\ and\ \citenamefont
  {Papoian}(2009{\natexlab{b}})}]{Savelyev:2009}%
  \BibitemOpen
  \bibfield  {author} {\bibinfo {author} {\bibfnamefont {A.}~\bibnamefont
  {Savelyev}}\ and\ \bibinfo {author} {\bibfnamefont {G.~A.}\ \bibnamefont
  {Papoian}},\ }\href {\doibase {10.1021/jp9005058}} {\bibfield  {journal}
  {\bibinfo  {journal} {J. Phys. Chem. B}\ }\textbf {\bibinfo {volume} {113}},\
  \bibinfo {pages} {7785} (\bibinfo {year} {2009}{\natexlab{b}})}\BibitemShut
  {NoStop}%
\bibitem [{\citenamefont {Moradzadeh}\ \emph {et~al.}(2018)\citenamefont
  {Moradzadeh}, \citenamefont {Motevaselian}, \citenamefont {Mashayak},\ and\
  \citenamefont {Aluru}}]{Moradzadeh:2018}%
  \BibitemOpen
  \bibfield  {author} {\bibinfo {author} {\bibfnamefont {A.}~\bibnamefont
  {Moradzadeh}}, \bibinfo {author} {\bibfnamefont {M.~H.}\ \bibnamefont
  {Motevaselian}}, \bibinfo {author} {\bibfnamefont {S.~Y.}\ \bibnamefont
  {Mashayak}}, \ and\ \bibinfo {author} {\bibfnamefont {N.~R.}\ \bibnamefont
  {Aluru}},\ }\href@noop {} {\bibfield  {journal} {\bibinfo  {journal} {J.
  Chem. Theor. Comp.}\ }\textbf {\bibinfo {volume} {14}},\ \bibinfo {pages}
  {3252} (\bibinfo {year} {2018})}\BibitemShut {NoStop}%
\bibitem [{\citenamefont {Murtola}, \citenamefont {Karttunen},\ and\
  \citenamefont {Vattulainen}(2009)}]{Murtola:2009ys}%
  \BibitemOpen
  \bibfield  {author} {\bibinfo {author} {\bibfnamefont {T.}~\bibnamefont
  {Murtola}}, \bibinfo {author} {\bibfnamefont {M.}~\bibnamefont {Karttunen}},
  \ and\ \bibinfo {author} {\bibfnamefont {I.}~\bibnamefont {Vattulainen}},\
  }\href {\doibase {10.1063/1.3167405}} {\bibfield  {journal} {\bibinfo
  {journal} {J. Chem. Phys.}\ }\textbf {\bibinfo {volume} {131}},\ \bibinfo
  {pages} {055101} (\bibinfo {year} {2009})}\BibitemShut {NoStop}%
\bibitem [{\citenamefont {Rudzinski}\ and\ \citenamefont
  {Noid}(2011)}]{Rudzinski:2011}%
  \BibitemOpen
  \bibfield  {author} {\bibinfo {author} {\bibfnamefont {J.~F.}\ \bibnamefont
  {Rudzinski}}\ and\ \bibinfo {author} {\bibfnamefont {W.~G.}\ \bibnamefont
  {Noid}},\ }\href {\doibase {10.1063/1.3663709}} {\bibfield  {journal}
  {\bibinfo  {journal} {J. Chem. Phys.}\ }\textbf {\bibinfo {volume} {135}},\
  \bibinfo {pages} {214101} (\bibinfo {year} {2011})}\BibitemShut {NoStop}%
\bibitem [{\citenamefont {Bilionis}\ and\ \citenamefont
  {Zabaras}(2013)}]{Bilionis:2013vn}%
  \BibitemOpen
  \bibfield  {author} {\bibinfo {author} {\bibfnamefont {I.}~\bibnamefont
  {Bilionis}}\ and\ \bibinfo {author} {\bibfnamefont {N.}~\bibnamefont
  {Zabaras}},\ }\href {\doibase 10.1063/1.4789308} {\bibfield  {journal}
  {\bibinfo  {journal} {J. Chem. Phys.}\ }\textbf {\bibinfo {volume} {138}},\
  \bibinfo {pages} {044313} (\bibinfo {year} {2013})}\BibitemShut {NoStop}%
\bibitem [{\citenamefont {Shell}(2016)}]{Shell:2016}%
  \BibitemOpen
  \bibfield  {author} {\bibinfo {author} {\bibfnamefont {M.~S.}\ \bibnamefont
  {Shell}},\ }in\ \href@noop {} {\emph {\bibinfo {booktitle} {Advances in
  Chemical Physics}}},\ Vol.\ \bibinfo {volume} {161},\ \bibinfo {editor}
  {edited by\ \bibinfo {editor} {\bibnamefont {{Rice, SA and Dinner, AR}}}}\
  (\bibinfo  {publisher} {Wiley-Blackwell},\ \bibinfo {address} {Malden, MA,
  USA},\ \bibinfo {year} {2016})\ pp.\ \bibinfo {pages} {395--441}\BibitemShut
  {NoStop}%
\bibitem [{\citenamefont {Jain}, \citenamefont {Garde},\ and\ \citenamefont
  {Kumar}(2006)}]{Jain:2006nx}%
  \BibitemOpen
  \bibfield  {author} {\bibinfo {author} {\bibfnamefont {S.}~\bibnamefont
  {Jain}}, \bibinfo {author} {\bibfnamefont {S.}~\bibnamefont {Garde}}, \ and\
  \bibinfo {author} {\bibfnamefont {S.~K.}\ \bibnamefont {Kumar}},\ }\href
  {\doibase 10.1021/ie060042h} {\bibfield  {journal} {\bibinfo  {journal} {Ind.
  Eng. Chem. Res.}\ }\textbf {\bibinfo {volume} {45}},\ \bibinfo {pages} {5614}
  (\bibinfo {year} {2006})}\BibitemShut {NoStop}%
\bibitem [{\citenamefont {Megariotis}\ \emph {et~al.}(2011)\citenamefont
  {Megariotis}, \citenamefont {Vyrkou}, \citenamefont {Leygue},\ and\
  \citenamefont {Theodorou}}]{Megariotis:2010fk}%
  \BibitemOpen
  \bibfield  {author} {\bibinfo {author} {\bibfnamefont {G.}~\bibnamefont
  {Megariotis}}, \bibinfo {author} {\bibfnamefont {A.}~\bibnamefont {Vyrkou}},
  \bibinfo {author} {\bibfnamefont {A.}~\bibnamefont {Leygue}}, \ and\ \bibinfo
  {author} {\bibfnamefont {D.~N.}\ \bibnamefont {Theodorou}},\ }\href {\doibase
  {10.1021/ie901957r}} {\bibfield  {journal} {\bibinfo  {journal} {Ind. Eng.
  Chem. Res.}\ }\textbf {\bibinfo {volume} {50}},\ \bibinfo {pages} {546}
  (\bibinfo {year} {2011})}\BibitemShut {NoStop}%
\bibitem [{\citenamefont {Fu}\ \emph {et~al.}(2012)\citenamefont {Fu},
  \citenamefont {Kulkarni}, \citenamefont {Shell},\ and\ \citenamefont
  {Leal}}]{Fu:2012oq}%
  \BibitemOpen
  \bibfield  {author} {\bibinfo {author} {\bibfnamefont {C.-C.}\ \bibnamefont
  {Fu}}, \bibinfo {author} {\bibfnamefont {P.~M.}\ \bibnamefont {Kulkarni}},
  \bibinfo {author} {\bibfnamefont {M.~S.}\ \bibnamefont {Shell}}, \ and\
  \bibinfo {author} {\bibfnamefont {L.~G.}\ \bibnamefont {Leal}},\ }\href
  {\doibase 10.1063/1.4759463} {\bibfield  {journal} {\bibinfo  {journal} {J.
  Chem. Phys.}\ }\textbf {\bibinfo {volume} {137}},\ \bibinfo {pages} {164106}
  (\bibinfo {year} {2012})}\BibitemShut {NoStop}%
\bibitem [{\citenamefont {Moore}, \citenamefont {Iacovella},\ and\
  \citenamefont {McCabe}(2014)}]{Moore:2014}%
  \BibitemOpen
  \bibfield  {author} {\bibinfo {author} {\bibfnamefont {T.~C.}\ \bibnamefont
  {Moore}}, \bibinfo {author} {\bibfnamefont {C.~R.}\ \bibnamefont
  {Iacovella}}, \ and\ \bibinfo {author} {\bibfnamefont {C.}~\bibnamefont
  {McCabe}},\ }\href@noop {} {\bibfield  {journal} {\bibinfo  {journal} {J.
  Chem. Phys.}\ }\textbf {\bibinfo {volume} {{140}}},\ \bibinfo {pages}
  {{224104}} (\bibinfo {year} {{2014}})}\BibitemShut {NoStop}%
\bibitem [{\citenamefont {Rosenberger}, \citenamefont {Hanke},\ and\
  \citenamefont {van~der Vegt}(2016)}]{Rosenberger:2016}%
  \BibitemOpen
  \bibfield  {author} {\bibinfo {author} {\bibfnamefont {D.}~\bibnamefont
  {Rosenberger}}, \bibinfo {author} {\bibfnamefont {M.}~\bibnamefont {Hanke}},
  \ and\ \bibinfo {author} {\bibfnamefont {N.~F.~A.}\ \bibnamefont {van~der
  Vegt}},\ }\href@noop {} {\bibfield  {journal} {\bibinfo  {journal} {Eur.
  Phys. J. Special Topics}\ }\textbf {\bibinfo {volume} {{225}}},\ \bibinfo
  {pages} {1323} (\bibinfo {year} {{2016}})}\BibitemShut {NoStop}%
\bibitem [{\citenamefont {Das}\ and\ \citenamefont
  {Andersen}(2010)}]{Das:2010}%
  \BibitemOpen
  \bibfield  {author} {\bibinfo {author} {\bibfnamefont {A.}~\bibnamefont
  {Das}}\ and\ \bibinfo {author} {\bibfnamefont {H.~C.}\ \bibnamefont
  {Andersen}},\ }\href {\doibase {10.1063/1.3394862}} {\bibfield  {journal}
  {\bibinfo  {journal} {J. Chem. Phys.}\ }\textbf {\bibinfo {volume} {132}},\
  \bibinfo {pages} {164106} (\bibinfo {year} {2010})}\BibitemShut {NoStop}%
\bibitem [{\citenamefont {Dunn}\ and\ \citenamefont {Noid}(2015)}]{Dunn:2015}%
  \BibitemOpen
  \bibfield  {author} {\bibinfo {author} {\bibfnamefont {N.~J.~H.}\
  \bibnamefont {Dunn}}\ and\ \bibinfo {author} {\bibfnamefont {W.~G.}\
  \bibnamefont {Noid}},\ }\href@noop {} {\bibfield  {journal} {\bibinfo
  {journal} {J. Chem. Phys.}\ }\textbf {\bibinfo {volume} {143}} (\bibinfo
  {year} {2015})}\BibitemShut {NoStop}%
\bibitem [{\citenamefont {Dunn}\ and\ \citenamefont {Noid}(2016)}]{Dunn:2016}%
  \BibitemOpen
  \bibfield  {author} {\bibinfo {author} {\bibfnamefont {N.~J.~H.}\
  \bibnamefont {Dunn}}\ and\ \bibinfo {author} {\bibfnamefont {W.~G.}\
  \bibnamefont {Noid}},\ }\href@noop {} {\bibfield  {journal} {\bibinfo
  {journal} {J. Chem. Phys.}\ }\textbf {\bibinfo {volume} {{144}}},\ \bibinfo
  {pages} {204124} (\bibinfo {year} {{2016}})}\BibitemShut {NoStop}%
\bibitem [{\citenamefont {Izvekov}\ and\ \citenamefont
  {Voth}(2006)}]{Izvekov:2006}%
  \BibitemOpen
  \bibfield  {author} {\bibinfo {author} {\bibfnamefont {S.}~\bibnamefont
  {Izvekov}}\ and\ \bibinfo {author} {\bibfnamefont {G.~A.}\ \bibnamefont
  {Voth}},\ }\href {\doibase {10.1063/1.2360580}} {\bibfield  {journal}
  {\bibinfo  {journal} {J. Chem. Phys.}\ }\textbf {\bibinfo {volume} {125}},\
  \bibinfo {pages} {151101} (\bibinfo {year} {2006})}\BibitemShut {NoStop}%
\bibitem [{\citenamefont {Hij\'{o}n}\ \emph {et~al.}(2010)\citenamefont
  {Hij\'{o}n}, \citenamefont {Espa\~{n}ol}, \citenamefont {Vanden-Eijnden},\
  and\ \citenamefont {Delgado-Buscalioni}}]{Hijon:2010}%
  \BibitemOpen
  \bibfield  {author} {\bibinfo {author} {\bibfnamefont {C.}~\bibnamefont
  {Hij\'{o}n}}, \bibinfo {author} {\bibfnamefont {P.}~\bibnamefont
  {Espa\~{n}ol}}, \bibinfo {author} {\bibfnamefont {E.}~\bibnamefont
  {Vanden-Eijnden}}, \ and\ \bibinfo {author} {\bibfnamefont {R.}~\bibnamefont
  {Delgado-Buscalioni}},\ }\href {\doibase 10.1039/b919800h} {\bibfield
  {journal} {\bibinfo  {journal} {Faraday Disc.}\ }\textbf {\bibinfo {volume}
  {144}},\ \bibinfo {pages} {301} (\bibinfo {year} {2010})}\BibitemShut
  {NoStop}%
\bibitem [{\citenamefont {Deichmann}\ and\ \citenamefont {van~der
  Vegt}(2018)}]{Deichmann:2018}%
  \BibitemOpen
  \bibfield  {author} {\bibinfo {author} {\bibfnamefont {G.}~\bibnamefont
  {Deichmann}}\ and\ \bibinfo {author} {\bibfnamefont {N.~F.~A.}\ \bibnamefont
  {van~der Vegt}},\ }\href@noop {} {\bibfield  {journal} {\bibinfo  {journal}
  {J. Chem. Phys.}\ }\textbf {\bibinfo {volume} {149}},\ \bibinfo {pages}
  {244114} (\bibinfo {year} {2018})}\BibitemShut {NoStop}%
\bibitem [{\citenamefont {Praprotnik}\ \emph {et~al.}(2007)\citenamefont
  {Praprotnik}, \citenamefont {Matysiak}, \citenamefont {Delle~Site},
  \citenamefont {Kremer},\ and\ \citenamefont {Clementi}}]{Praprotnik:2007bl}%
  \BibitemOpen
  \bibfield  {author} {\bibinfo {author} {\bibfnamefont {M.}~\bibnamefont
  {Praprotnik}}, \bibinfo {author} {\bibfnamefont {S.}~\bibnamefont
  {Matysiak}}, \bibinfo {author} {\bibfnamefont {L.}~\bibnamefont
  {Delle~Site}}, \bibinfo {author} {\bibfnamefont {K.}~\bibnamefont {Kremer}},
  \ and\ \bibinfo {author} {\bibfnamefont {C.}~\bibnamefont {Clementi}},\
  }\href@noop {} {\bibfield  {journal} {\bibinfo  {journal} {J. Phys.: Condens.
  Matter}\ }\textbf {\bibinfo {volume} {19}} (\bibinfo {year}
  {2007})}\BibitemShut {NoStop}%
\bibitem [{\citenamefont {Wang}, \citenamefont {Junghans},\ and\ \citenamefont
  {Kremer}(2009)}]{Wang:2009ol}%
  \BibitemOpen
  \bibfield  {author} {\bibinfo {author} {\bibfnamefont {H.}~\bibnamefont
  {Wang}}, \bibinfo {author} {\bibfnamefont {C.}~\bibnamefont {Junghans}}, \
  and\ \bibinfo {author} {\bibfnamefont {K.}~\bibnamefont {Kremer}},\ }\href
  {\doibase {10.1140/epje/i2008-10413-5}} {\bibfield  {journal} {\bibinfo
  {journal} {Eur. Phys. J. E}\ }\textbf {\bibinfo {volume} {28}},\ \bibinfo
  {pages} {221} (\bibinfo {year} {2009})}\BibitemShut {NoStop}%
\bibitem [{\citenamefont {Molinero}\ and\ \citenamefont
  {Moore}(2009)}]{Molinero:2009fu}%
  \BibitemOpen
  \bibfield  {author} {\bibinfo {author} {\bibfnamefont {V.}~\bibnamefont
  {Molinero}}\ and\ \bibinfo {author} {\bibfnamefont {E.~B.}\ \bibnamefont
  {Moore}},\ }\href {\doibase 10.1021/jp805227c} {\bibfield  {journal}
  {\bibinfo  {journal} {J. Phys. Chem. B}\ }\textbf {\bibinfo {volume} {113}},\
  \bibinfo {pages} {4008} (\bibinfo {year} {2009})}\BibitemShut {NoStop}%
\bibitem [{\citenamefont {Jochum}\ \emph {et~al.}(2012)\citenamefont {Jochum},
  \citenamefont {Andrienko}, \citenamefont {Kremer},\ and\ \citenamefont
  {Peter}}]{Jochum:2012wa}%
  \BibitemOpen
  \bibfield  {author} {\bibinfo {author} {\bibfnamefont {M.}~\bibnamefont
  {Jochum}}, \bibinfo {author} {\bibfnamefont {D.}~\bibnamefont {Andrienko}},
  \bibinfo {author} {\bibfnamefont {K.}~\bibnamefont {Kremer}}, \ and\ \bibinfo
  {author} {\bibfnamefont {C.}~\bibnamefont {Peter}},\ }\href {\doibase
  10.1063/1.4742067} {\bibfield  {journal} {\bibinfo  {journal} {J. Chem.
  Phys.}\ }\textbf {\bibinfo {volume} {137}},\ \bibinfo {pages} {064102}
  (\bibinfo {year} {2012})}\BibitemShut {NoStop}%
\bibitem [{\citenamefont {Fritsch}\ \emph {et~al.}(2014)\citenamefont
  {Fritsch}, \citenamefont {Potestio}, \citenamefont {Donadio},\ and\
  \citenamefont {Kremer}}]{Fritsch:2014}%
  \BibitemOpen
  \bibfield  {author} {\bibinfo {author} {\bibfnamefont {S.}~\bibnamefont
  {Fritsch}}, \bibinfo {author} {\bibfnamefont {R.}~\bibnamefont {Potestio}},
  \bibinfo {author} {\bibfnamefont {D.}~\bibnamefont {Donadio}}, \ and\
  \bibinfo {author} {\bibfnamefont {K.}~\bibnamefont {Kremer}},\ }\href@noop {}
  {\bibfield  {journal} {\bibinfo  {journal} {J. Chem. Theor. Comp.}\ }\textbf
  {\bibinfo {volume} {{10}}},\ \bibinfo {pages} {816} (\bibinfo {year}
  {{2014}})}\BibitemShut {NoStop}%
\bibitem [{\citenamefont {Hadley}\ and\ \citenamefont
  {McCabe}(2012)}]{Hadley:2012}%
  \BibitemOpen
  \bibfield  {author} {\bibinfo {author} {\bibfnamefont {K.~R.}\ \bibnamefont
  {Hadley}}\ and\ \bibinfo {author} {\bibfnamefont {C.}~\bibnamefont
  {McCabe}},\ }\href@noop {} {\bibfield  {journal} {\bibinfo  {journal} {Mol.
  Simulat.}\ }\textbf {\bibinfo {volume} {{38}}},\ \bibinfo {pages} {671}
  (\bibinfo {year} {{2012}})}\BibitemShut {NoStop}%
\bibitem [{\citenamefont {R{\"u}hle}\ \emph {et~al.}(2009)\citenamefont
  {R{\"u}hle}, \citenamefont {Junghans}, \citenamefont {Lukyanov},
  \citenamefont {Kremer},\ and\ \citenamefont {Andrienko}}]{Ruhle:2009wx}%
  \BibitemOpen
  \bibfield  {author} {\bibinfo {author} {\bibfnamefont {V.}~\bibnamefont
  {R{\"u}hle}}, \bibinfo {author} {\bibfnamefont {C.}~\bibnamefont {Junghans}},
  \bibinfo {author} {\bibfnamefont {A.}~\bibnamefont {Lukyanov}}, \bibinfo
  {author} {\bibfnamefont {K.}~\bibnamefont {Kremer}}, \ and\ \bibinfo {author}
  {\bibfnamefont {D.}~\bibnamefont {Andrienko}},\ }\href {\doibase
  {10.1021/ct900369w}} {\bibfield  {journal} {\bibinfo  {journal} {J. Chem.
  Theor. Comp.}\ }\textbf {\bibinfo {volume} {5}},\ \bibinfo {pages} {3211}
  (\bibinfo {year} {2009})}\BibitemShut {NoStop}%
\bibitem [{\citenamefont {Chaimovich}\ and\ \citenamefont
  {Shell}(2009)}]{Chaimovich:2009tw}%
  \BibitemOpen
  \bibfield  {author} {\bibinfo {author} {\bibfnamefont {A.}~\bibnamefont
  {Chaimovich}}\ and\ \bibinfo {author} {\bibfnamefont {M.~S.}\ \bibnamefont
  {Shell}},\ }\href {\doibase 10.1039/b818512c} {\bibfield  {journal} {\bibinfo
   {journal} {Phys. Chem. Chem. Phys.}\ }\textbf {\bibinfo {volume} {11}},\
  \bibinfo {pages} {1901} (\bibinfo {year} {2009})}\BibitemShut {NoStop}%
\bibitem [{\citenamefont {Larini}, \citenamefont {Lu},\ and\ \citenamefont
  {Voth}(2010)}]{Larini:2010dq}%
  \BibitemOpen
  \bibfield  {author} {\bibinfo {author} {\bibfnamefont {L.}~\bibnamefont
  {Larini}}, \bibinfo {author} {\bibfnamefont {L.~Y.}\ \bibnamefont {Lu}}, \
  and\ \bibinfo {author} {\bibfnamefont {G.~A.}\ \bibnamefont {Voth}},\ }\href
  {\doibase {10.1063/1.3394863}} {\bibfield  {journal} {\bibinfo  {journal} {J.
  Chem. Phys.}\ }\textbf {\bibinfo {volume} {132}},\ \bibinfo {pages} {164107}
  (\bibinfo {year} {2010})}\BibitemShut {NoStop}%
\bibitem [{\citenamefont {Das}\ and\ \citenamefont
  {Andersen}(2012)}]{Das:2012c}%
  \BibitemOpen
  \bibfield  {author} {\bibinfo {author} {\bibfnamefont {A.}~\bibnamefont
  {Das}}\ and\ \bibinfo {author} {\bibfnamefont {H.~C.}\ \bibnamefont
  {Andersen}},\ }\href {\doibase {10.1063/1.4705417}} {\bibfield  {journal}
  {\bibinfo  {journal} {J. Chem. Phys.}\ }\textbf {\bibinfo {volume} {{136}}},\
  \bibinfo {pages} {{194114}} (\bibinfo {year} {{2012}})}\BibitemShut {NoStop}%
\bibitem [{\citenamefont {Lu}\ \emph {et~al.}(2014)\citenamefont {Lu},
  \citenamefont {Qiu}, \citenamefont {Baron},\ and\ \citenamefont
  {Molinero}}]{Lu:2014jctc}%
  \BibitemOpen
  \bibfield  {author} {\bibinfo {author} {\bibfnamefont {J.}~\bibnamefont
  {Lu}}, \bibinfo {author} {\bibfnamefont {Y.}~\bibnamefont {Qiu}}, \bibinfo
  {author} {\bibfnamefont {R.}~\bibnamefont {Baron}}, \ and\ \bibinfo {author}
  {\bibfnamefont {V.}~\bibnamefont {Molinero}},\ }\href@noop {} {\bibfield
  {journal} {\bibinfo  {journal} {J. Chem. Theor. Comp.}\ }\textbf {\bibinfo
  {volume} {{10}}},\ \bibinfo {pages} {4104} (\bibinfo {year}
  {{2014}})}\BibitemShut {NoStop}%
\bibitem [{\citenamefont {Scherer}\ and\ \citenamefont
  {Andrienko}(2018)}]{Scherer:2018}%
  \BibitemOpen
  \bibfield  {author} {\bibinfo {author} {\bibfnamefont {C.}~\bibnamefont
  {Scherer}}\ and\ \bibinfo {author} {\bibfnamefont {D.}~\bibnamefont
  {Andrienko}},\ }\href@noop {} {\bibfield  {journal} {\bibinfo  {journal}
  {Phys. Chem. Chem. Phys.}\ }\textbf {\bibinfo {volume} {20}},\ \bibinfo
  {pages} {22387} (\bibinfo {year} {2018})}\BibitemShut {NoStop}%
\bibitem [{\citenamefont {Weeks}, \citenamefont {Chandler},\ and\ \citenamefont
  {Andersen}(1971)}]{Weeks:1971}%
  \BibitemOpen
  \bibfield  {author} {\bibinfo {author} {\bibfnamefont {J.}~\bibnamefont
  {Weeks}}, \bibinfo {author} {\bibfnamefont {D.}~\bibnamefont {Chandler}}, \
  and\ \bibinfo {author} {\bibfnamefont {H.}~\bibnamefont {Andersen}},\ }\href
  {\doibase {10.1063/1.1674820}} {\bibfield  {journal} {\bibinfo  {journal} {J.
  Chem. Phys.}\ }\textbf {\bibinfo {volume} {{54}}},\ \bibinfo {pages} {5237}
  (\bibinfo {year} {{1971}})}\BibitemShut {NoStop}%
\bibitem [{\citenamefont {Yan}\ \emph {et~al.}(2006)\citenamefont {Yan},
  \citenamefont {Buldyrev}, \citenamefont {Giovambattista}, \citenamefont
  {Debenedetti},\ and\ \citenamefont {Stanley}}]{Yan:2006}%
  \BibitemOpen
  \bibfield  {author} {\bibinfo {author} {\bibfnamefont {Z.}~\bibnamefont
  {Yan}}, \bibinfo {author} {\bibfnamefont {S.~V.}\ \bibnamefont {Buldyrev}},
  \bibinfo {author} {\bibfnamefont {N.}~\bibnamefont {Giovambattista}},
  \bibinfo {author} {\bibfnamefont {P.~G.}\ \bibnamefont {Debenedetti}}, \ and\
  \bibinfo {author} {\bibfnamefont {H.~E.}\ \bibnamefont {Stanley}},\
  }\href@noop {} {\bibfield  {journal} {\bibinfo  {journal} {Phys. Rev. E}\
  }\textbf {\bibinfo {volume} {{73}}},\ \bibinfo {pages} {051204} (\bibinfo
  {year} {{2006}})}\BibitemShut {NoStop}%
\bibitem [{\citenamefont {Yan}\ \emph {et~al.}(2008)\citenamefont {Yan},
  \citenamefont {Buldyrev}, \citenamefont {Kumar}, \citenamefont
  {Giovambattista},\ and\ \citenamefont {Stanley}}]{Yan:2008}%
  \BibitemOpen
  \bibfield  {author} {\bibinfo {author} {\bibfnamefont {Z.}~\bibnamefont
  {Yan}}, \bibinfo {author} {\bibfnamefont {S.~V.}\ \bibnamefont {Buldyrev}},
  \bibinfo {author} {\bibfnamefont {P.}~\bibnamefont {Kumar}}, \bibinfo
  {author} {\bibfnamefont {N.}~\bibnamefont {Giovambattista}}, \ and\ \bibinfo
  {author} {\bibfnamefont {H.~E.}\ \bibnamefont {Stanley}},\ }\href@noop {}
  {\bibfield  {journal} {\bibinfo  {journal} {Phys. Rev. E}\ }\textbf {\bibinfo
  {volume} {77}},\ \bibinfo {pages} {042201} (\bibinfo {year}
  {{2008}})}\BibitemShut {NoStop}%
\bibitem [{\citenamefont {Noid}\ \emph
  {et~al.}(2008{\natexlab{b}})\citenamefont {Noid}, \citenamefont {Liu},
  \citenamefont {Wang}, \citenamefont {Chu}, \citenamefont {Ayton},
  \citenamefont {Izvekov}, \citenamefont {Andersen},\ and\ \citenamefont
  {Voth}}]{Noid:2008b}%
  \BibitemOpen
  \bibfield  {author} {\bibinfo {author} {\bibfnamefont {W.~G.}\ \bibnamefont
  {Noid}}, \bibinfo {author} {\bibfnamefont {P.}~\bibnamefont {Liu}}, \bibinfo
  {author} {\bibfnamefont {Y.~T.}\ \bibnamefont {Wang}}, \bibinfo {author}
  {\bibfnamefont {J.-W.}\ \bibnamefont {Chu}}, \bibinfo {author} {\bibfnamefont
  {G.~S.}\ \bibnamefont {Ayton}}, \bibinfo {author} {\bibfnamefont
  {S.}~\bibnamefont {Izvekov}}, \bibinfo {author} {\bibfnamefont {H.~C.}\
  \bibnamefont {Andersen}}, \ and\ \bibinfo {author} {\bibfnamefont {G.~A.}\
  \bibnamefont {Voth}},\ }\href {\doibase {10.1063/1.2938857}} {\bibfield
  {journal} {\bibinfo  {journal} {J. Chem. Phys.}\ }\textbf {\bibinfo {volume}
  {128}},\ \bibinfo {pages} {244115} (\bibinfo {year}
  {2008}{\natexlab{b}})}\BibitemShut {NoStop}%
\bibitem [{\citenamefont {Rudzinski}\ and\ \citenamefont
  {Noid}(2015{\natexlab{b}})}]{Rudzinski:2014b}%
  \BibitemOpen
  \bibfield  {author} {\bibinfo {author} {\bibfnamefont {J.~F.}\ \bibnamefont
  {Rudzinski}}\ and\ \bibinfo {author} {\bibfnamefont {W.~G.}\ \bibnamefont
  {Noid}},\ }\href {\doibase {10.1021/ct5009922}} {\bibfield  {journal}
  {\bibinfo  {journal} {J. Chem. Theor. Comp.}\ }\textbf {\bibinfo {volume}
  {{11}}},\ \bibinfo {pages} {1278} (\bibinfo {year}
  {2015}{\natexlab{b}})}\BibitemShut {NoStop}%
\bibitem [{\citenamefont {Suhubi}(2003)}]{Suhubi:2003}%
  \BibitemOpen
  \bibfield  {author} {\bibinfo {author} {\bibfnamefont {E.~S.}\ \bibnamefont
  {Suhubi}},\ }\href {\doibase https://doi.org/10.1007/978-94-017-0141-9}
  {\emph {\bibinfo {title} {Functional Analysis}}}\ (\bibinfo  {publisher}
  {Springer Dordrecht},\ \bibinfo {year} {2003})\ \bibinfo {note} {pages
  390-392}\BibitemShut {NoStop}%
\bibitem [{\citenamefont {Berendsen}, \citenamefont {Grigera},\ and\
  \citenamefont {Straatsma}(1987)}]{Berendsen:1987}%
  \BibitemOpen
  \bibfield  {author} {\bibinfo {author} {\bibfnamefont {H.~J.~C.}\
  \bibnamefont {Berendsen}}, \bibinfo {author} {\bibfnamefont {J.}~\bibnamefont
  {Grigera}}, \ and\ \bibinfo {author} {\bibfnamefont {T.~P.}\ \bibnamefont
  {Straatsma}},\ }\href {\doibase 10.1021/j100308a038} {\bibfield  {journal}
  {\bibinfo  {journal} {The Journal of Physical Chemistry}\ }\textbf {\bibinfo
  {volume} {91}},\ \bibinfo {pages} {6269} (\bibinfo {year}
  {1987})}\BibitemShut {NoStop}%
\bibitem [{\citenamefont {Abraham}\ \emph {et~al.}(2015)\citenamefont
  {Abraham}, \citenamefont {Murtola}, \citenamefont {Schulz}, \citenamefont
  {Páll}, \citenamefont {Smith}, \citenamefont {Hess},\ and\ \citenamefont
  {Lindahl}}]{Abraham:2015}%
  \BibitemOpen
  \bibfield  {author} {\bibinfo {author} {\bibfnamefont {M.~J.}\ \bibnamefont
  {Abraham}}, \bibinfo {author} {\bibfnamefont {T.}~\bibnamefont {Murtola}},
  \bibinfo {author} {\bibfnamefont {R.}~\bibnamefont {Schulz}}, \bibinfo
  {author} {\bibfnamefont {S.}~\bibnamefont {Páll}}, \bibinfo {author}
  {\bibfnamefont {J.~C.}\ \bibnamefont {Smith}}, \bibinfo {author}
  {\bibfnamefont {B.}~\bibnamefont {Hess}}, \ and\ \bibinfo {author}
  {\bibfnamefont {E.}~\bibnamefont {Lindahl}},\ }\href {\doibase
  https://doi.org/10.1016/j.softx.2015.06.001} {\bibfield  {journal} {\bibinfo
  {journal} {SoftwareX}\ }\textbf {\bibinfo {volume} {1-2}},\ \bibinfo {pages}
  {19 } (\bibinfo {year} {2015})}\BibitemShut {NoStop}%
\bibitem [{\citenamefont {Berendsen}\ \emph {et~al.}(1984)\citenamefont
  {Berendsen}, \citenamefont {Postma}, \citenamefont {van Gunsteren},
  \citenamefont {DiNola},\ and\ \citenamefont {Haak}}]{Berendsen:1984}%
  \BibitemOpen
  \bibfield  {author} {\bibinfo {author} {\bibfnamefont {H.~J.~C.}\
  \bibnamefont {Berendsen}}, \bibinfo {author} {\bibfnamefont {J.~P.~M.}\
  \bibnamefont {Postma}}, \bibinfo {author} {\bibfnamefont {W.~F.}\
  \bibnamefont {van Gunsteren}}, \bibinfo {author} {\bibfnamefont
  {A.}~\bibnamefont {DiNola}}, \ and\ \bibinfo {author} {\bibfnamefont {J.~R.}\
  \bibnamefont {Haak}},\ }\href {\doibase {10.1063/1.448118}} {\bibfield
  {journal} {\bibinfo  {journal} {J. Chem. Phys.}\ }\textbf {\bibinfo {volume}
  {81}},\ \bibinfo {pages} {3684} (\bibinfo {year} {1984})}\BibitemShut
  {NoStop}%
\bibitem [{\citenamefont {Gunsteren}\ and\ \citenamefont
  {Berendsen}(1988)}]{VanGunsteren:1988}%
  \BibitemOpen
  \bibfield  {author} {\bibinfo {author} {\bibfnamefont {W.~F.~V.}\
  \bibnamefont {Gunsteren}}\ and\ \bibinfo {author} {\bibfnamefont {H.~J.~C.}\
  \bibnamefont {Berendsen}},\ }\href {\doibase 10.1080/08927028808080941}
  {\bibfield  {journal} {\bibinfo  {journal} {Molecular Simulation}\ }\textbf
  {\bibinfo {volume} {1}},\ \bibinfo {pages} {173} (\bibinfo {year}
  {1988})}\BibitemShut {NoStop}%
\bibitem [{\citenamefont {Essmann}\ \emph {et~al.}(1995)\citenamefont
  {Essmann}, \citenamefont {Perera}, \citenamefont {Berkowitz}, \citenamefont
  {Darden}, \citenamefont {Lee},\ and\ \citenamefont
  {Pedersen}}]{Essmann:1995}%
  \BibitemOpen
  \bibfield  {author} {\bibinfo {author} {\bibfnamefont {U.}~\bibnamefont
  {Essmann}}, \bibinfo {author} {\bibfnamefont {L.}~\bibnamefont {Perera}},
  \bibinfo {author} {\bibfnamefont {M.~L.}\ \bibnamefont {Berkowitz}}, \bibinfo
  {author} {\bibfnamefont {T.}~\bibnamefont {Darden}}, \bibinfo {author}
  {\bibfnamefont {H.}~\bibnamefont {Lee}}, \ and\ \bibinfo {author}
  {\bibfnamefont {L.~G.}\ \bibnamefont {Pedersen}},\ }\href {\doibase
  10.1063/1.470117} {\bibfield  {journal} {\bibinfo  {journal} {The Journal of
  Chemical Physics}\ }\textbf {\bibinfo {volume} {103}},\ \bibinfo {pages}
  {8577} (\bibinfo {year} {1995})}\BibitemShut {NoStop}%
\bibitem [{\citenamefont {Mashayak}\ \emph {et~al.}(2015)\citenamefont
  {Mashayak}, \citenamefont {Jochum}, \citenamefont {Koschke}, \citenamefont
  {Aluru}, \citenamefont {Ruehle},\ and\ \citenamefont
  {Junghans}}]{Mashayak:2015}%
  \BibitemOpen
  \bibfield  {author} {\bibinfo {author} {\bibfnamefont {S.~Y.}\ \bibnamefont
  {Mashayak}}, \bibinfo {author} {\bibfnamefont {M.~N.}\ \bibnamefont
  {Jochum}}, \bibinfo {author} {\bibfnamefont {K.}~\bibnamefont {Koschke}},
  \bibinfo {author} {\bibfnamefont {N.~R.}\ \bibnamefont {Aluru}}, \bibinfo
  {author} {\bibfnamefont {V.}~\bibnamefont {Ruehle}}, \ and\ \bibinfo {author}
  {\bibfnamefont {C.}~\bibnamefont {Junghans}},\ }\href@noop {} {\bibfield
  {journal} {\bibinfo  {journal} {PLoS ONE}\ }\textbf {\bibinfo {volume}
  {10}},\ \bibinfo {pages} {1} (\bibinfo {year} {2015})}\BibitemShut {NoStop}%
\bibitem [{\citenamefont {Dunn}\ \emph {et~al.}(2018)\citenamefont {Dunn},
  \citenamefont {Lebold}, \citenamefont {DeLyser}, \citenamefont {Rudzinski},\
  and\ \citenamefont {Noid}}]{Dunn:2018}%
  \BibitemOpen
  \bibfield  {author} {\bibinfo {author} {\bibfnamefont {N.~J.~H.}\
  \bibnamefont {Dunn}}, \bibinfo {author} {\bibfnamefont {K.~M.}\ \bibnamefont
  {Lebold}}, \bibinfo {author} {\bibfnamefont {M.~R.}\ \bibnamefont {DeLyser}},
  \bibinfo {author} {\bibfnamefont {J.~F.}\ \bibnamefont {Rudzinski}}, \ and\
  \bibinfo {author} {\bibfnamefont {W.}~\bibnamefont {Noid}},\ }\href@noop {}
  {\bibfield  {journal} {\bibinfo  {journal} {J. Phys. Chem. B}\ }\textbf
  {\bibinfo {volume} {122}},\ \bibinfo {pages} {3363} (\bibinfo {year}
  {2018})}\BibitemShut {NoStop}%
\bibitem [{\citenamefont {Hess}\ \emph {et~al.}(2008)\citenamefont {Hess},
  \citenamefont {Kutzner}, \citenamefont {van~der Spoel},\ and\ \citenamefont
  {Lindahl}}]{Hess:2008}%
  \BibitemOpen
  \bibfield  {author} {\bibinfo {author} {\bibfnamefont {B.}~\bibnamefont
  {Hess}}, \bibinfo {author} {\bibfnamefont {C.}~\bibnamefont {Kutzner}},
  \bibinfo {author} {\bibfnamefont {D.}~\bibnamefont {van~der Spoel}}, \ and\
  \bibinfo {author} {\bibfnamefont {E.}~\bibnamefont {Lindahl}},\ }\href
  {\doibase {10.1021/ct700301q}} {\bibfield  {journal} {\bibinfo  {journal} {J.
  Chem. Theor. Comp.}\ }\textbf {\bibinfo {volume} {4}},\ \bibinfo {pages}
  {435} (\bibinfo {year} {2008})}\BibitemShut {NoStop}%
\bibitem [{\citenamefont {Jones}\ \emph {et~al.}(01  )\citenamefont {Jones},
  \citenamefont {Oliphant}, \citenamefont {Peterson} \emph {et~al.}}]{SciPy}%
  \BibitemOpen
  \bibfield  {author} {\bibinfo {author} {\bibfnamefont {E.}~\bibnamefont
  {Jones}}, \bibinfo {author} {\bibfnamefont {T.}~\bibnamefont {Oliphant}},
  \bibinfo {author} {\bibfnamefont {P.}~\bibnamefont {Peterson}},  \emph
  {et~al.},\ }\href {http://www.scipy.org/} {\enquote {\bibinfo {title}
  {{SciPy}: Open source scientific tools for {Python}},}\ } (\bibinfo {year}
  {2001--}),\ \bibinfo {note} {[Online; accessed May 2019]}\BibitemShut
  {NoStop}%
\bibitem [{\citenamefont {Bereau}\ and\ \citenamefont
  {Rudzinski}(2018)}]{Bereau:2018}%
  \BibitemOpen
  \bibfield  {author} {\bibinfo {author} {\bibfnamefont {T.}~\bibnamefont
  {Bereau}}\ and\ \bibinfo {author} {\bibfnamefont {J.~F.}\ \bibnamefont
  {Rudzinski}},\ }\href@noop {} {\bibfield  {journal} {\bibinfo  {journal}
  {Phys. Rev. Lett.}\ }\textbf {\bibinfo {volume} {121}},\ \bibinfo {pages}
  {256002} (\bibinfo {year} {2018})}\BibitemShut {NoStop}%
\end{thebibliography}%

\end{document}